*Review*

# Massive Star Formation in the Ultraviolet Observed with the Hubble Space Telescope

**Claus Leitherer**

Space Telescope Science Institute, 3700 San Martin Dr., Baltimore, MD 21218, USA; leitherer@stsci.edu

**Abstract:** Spectroscopic observations of a massive star formation in the ultraviolet and their interpretation are reviewed. After a brief historical retrospective, two well-studied resolved star clusters and the surrounding H II regions are introduced: NGC 2070 in the Large Magellanic Cloud and NGC 604 in M33. These regions serve as a training set for studies of more distant clusters, which can no longer be resolved into individual stars. Observations of recently formed star clusters and extended regions in star-forming galaxies in the nearby universe beyond the Local Group are presented. Their interpretation relies on spectral synthesis models. The successes and failures of such models are discussed, and future directions are highlighted. I present a case study of the extraordinary star cluster and giant H II region in the blue compact galaxy II Zw 40. The review concludes with a preview of two upcoming Hubble Space Telescope programs: ULLYSES, a survey of massive stars in nearby galaxies, and CLASSY, a study of massive star clusters in star-forming galaxies.

**Keywords:** star formation; ultraviolet; early-type stars; stellar evolution; star clusters; H II regions; late-type galaxies; starburst galaxies

---

## 1. Introduction—A Brief Retrospective of the Pre-Hubble Times

The space-ultraviolet (UV) wavelength region became accessible to spectroscopic observations in the 1960s. At this time, the introduction of the 3-axis star-pointing stabilization system on Aerobee sounding rockets enabled the acquisition of sufficiently deep spectrograms of astronomical objects outside the solar system [1]. These observations were restricted to bright stars; UV spectra of extragalactic star clusters and star-forming galaxies could not be collected until the launch of the International Ultraviolet Explorer (*IUE*) satellite, which had the capability of obtaining multi-hour exposures necessary for extragalactic studies [2]. Luminous young star clusters in Local Group galaxies such as the Large Magellanic Cloud (LMC) and M33 became preferred objects of study with *IUE*. In Figure 1, we show the UV spectra of two such clusters. NGC 2070 in the LMC is the ionizing source of the 30 Doradus nebula, which is the most luminous giant H II region in the Local Group [3]. Due to its proximity, the size of the entire cluster exceeds the aperture size of *IUE*, and mosaicking has been applied to construct a spectrum of the entire region [4]. In contrast, a single *IUE* spectrum of NGC 604 in M33 (right graphic in Figure 1) covers the entire cluster [5]. The UV spectra of both clusters are remarkably similar, displaying the characteristic spectral lines of massive stars, such as N V 1240, Si IV 1400, and C IV 1550.

The *IUE* satellite permitted extension of such studies to galaxies outside the Local Group. An atlas of all scientifically useful spectra of star-forming and active galaxies summarizes *IUE*'s data collection [2]. Figure 2 shows an example from this work. At a distance of 5.1 Mpc, NGC 1705 is one of the UV-brightest star-forming galaxies due to the presence of a single extremely bright star cluster [6]. Many of the spectral lines observed in NGC 2070 and NGC 604 are detected in NGC 1705 as well. The relatively high signal-to-noise of these spectra motivated quantitative studies to probe stellar population. The ratio of the Si IV 1400 and C IV 1550 stellar-wind lines can be utilized as a probe of

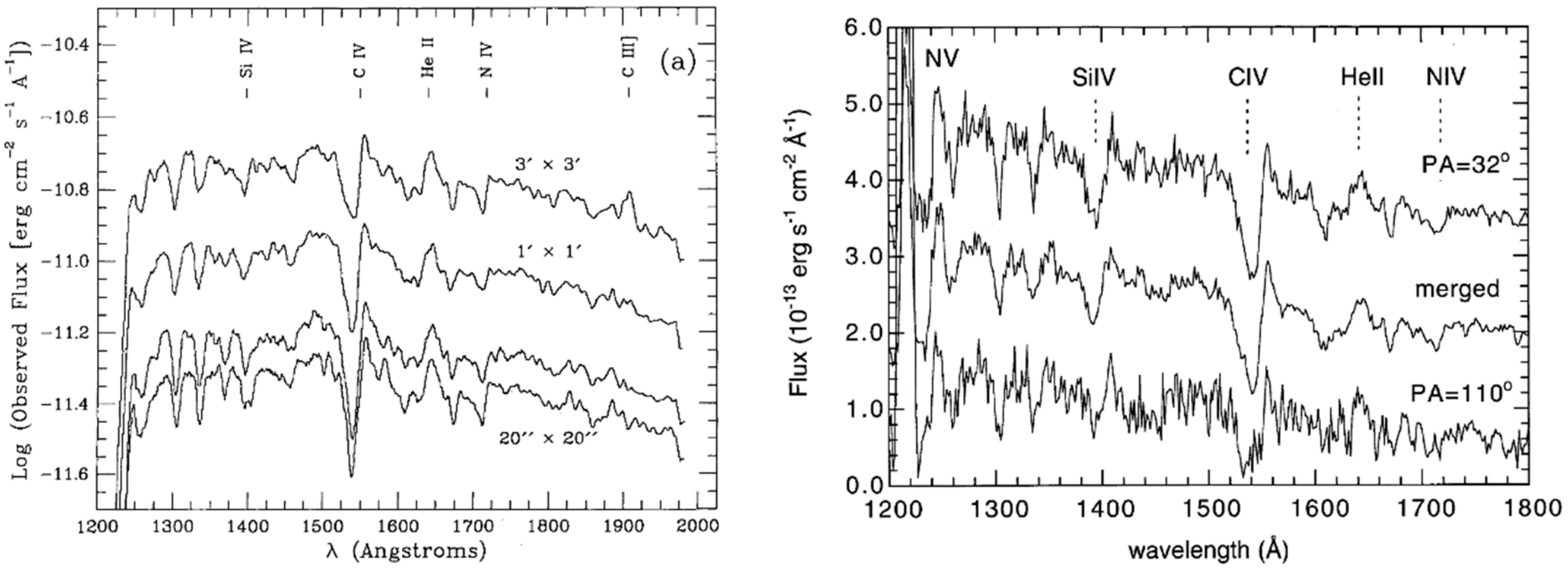

**Figure 1.** International Ultraviolet Explorer (*IUE*) spectra of NGC 2070 (**left**) and NGC 604 (**right**), two extragalactic massive star clusters and the Large Magellanic Cloud (LMC) and M33, respectively. The spectra show the typical UV spectral features of hot, massive stars. From [4,5].

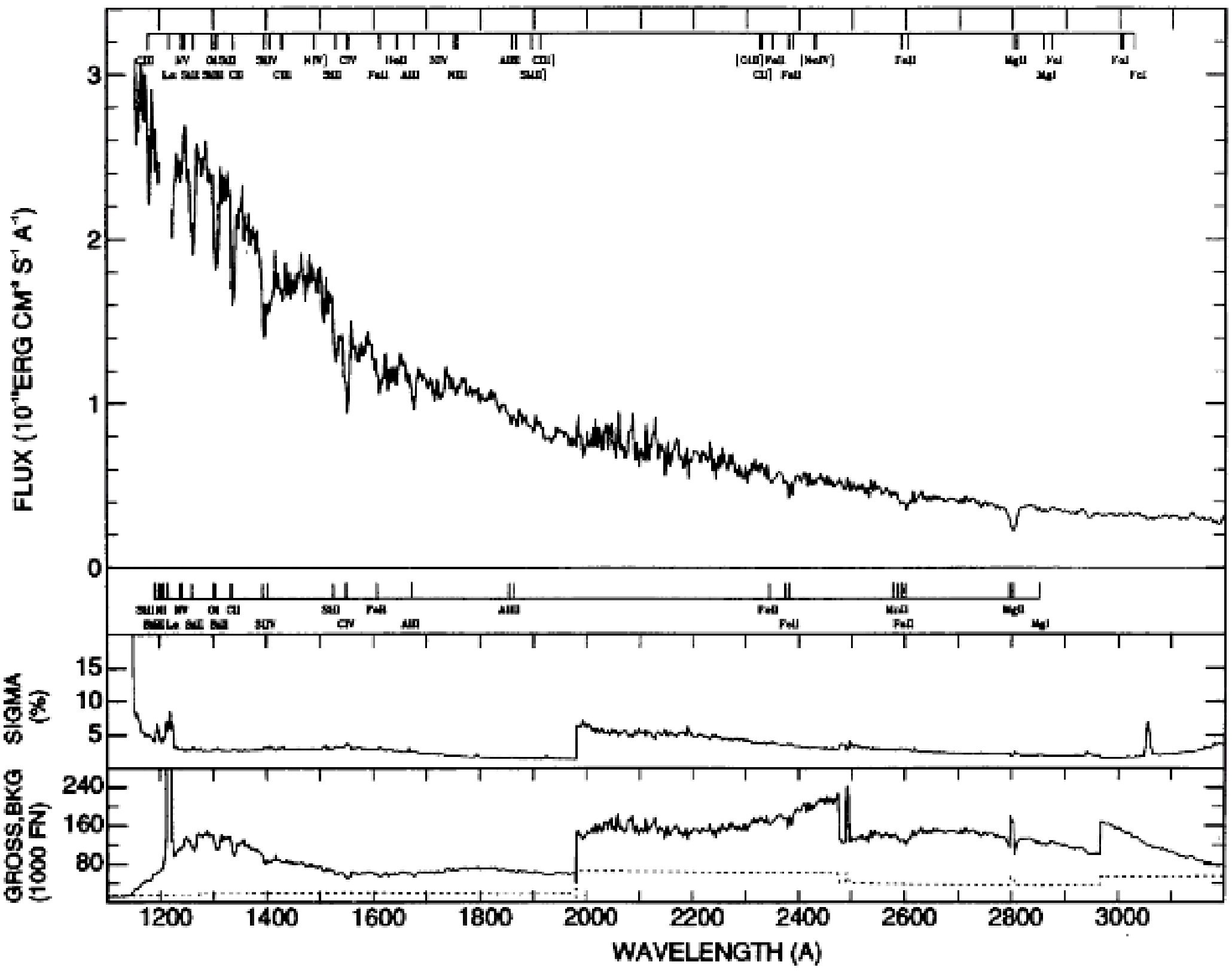

**Figure 2.** *IUE* low-dispersion spectrum of the nearby starburst galaxy NGC 1705, whose UV light is dominated by a bright super star cluster. The main observed and expected spectral features are identified at the top. The Milky Way foreground lines are labeled at the bottom. From [2].

The stellar initial mass function (IMF) [7]. The underlying physical reason is the different ionization energy of the two lines (33 eV versus 49 eV for Si IV and C IV, respectively), which affects the line ratio for different choices of the IMF. However, the strong P Cygni lines in stellar winds are mostly resonance transitions, which implies that they are strong in the interstellar medium (ISM) as well and can contribute to the features seen in Figure 2 [8]. The low spectral resolution of *IUE* low-dispersion spectra (~6 Å) is insufficient to separate the stellar and interstellar contributions. This concern turned out to be valid when spectra were obtained with the Hubble Space Telescope (*HST*). In Figure 3, we reproduced a spectrum of the same star cluster in NGC 1705 as the one in Figure 3 but obtained with *HST*'s Space Telescope Imaging Spectrograph (*STIS*) [6]. The broad, apparently single lines in the *IUE* spectrum now split into multiple components, and line features like N V 1240, Si IV 1400, and C IV 1550 are blends of narrow interstellar and broad stellar lines. Disentangling the stellar and interstellar lines requires a minimum spectral resolution of at least 1—2 Å, a requirement only met by the *HST*. The *HST*, together with the Hopkins Ultraviolet Telescope (*HUT*) and the Far-Ultraviolet Spectral Explorer (*FUSE*), led to order-of-magnitude improvements in the spectral resolution and the signal-to-noise over *IUE*. As I will discuss in the following sections, the wavelength region between 1200 and 2000 Å holds the most diagnostic power for spectral diagnostics. Therefore, I will mostly focus on data obtained with the *HST* in this wavelength region.

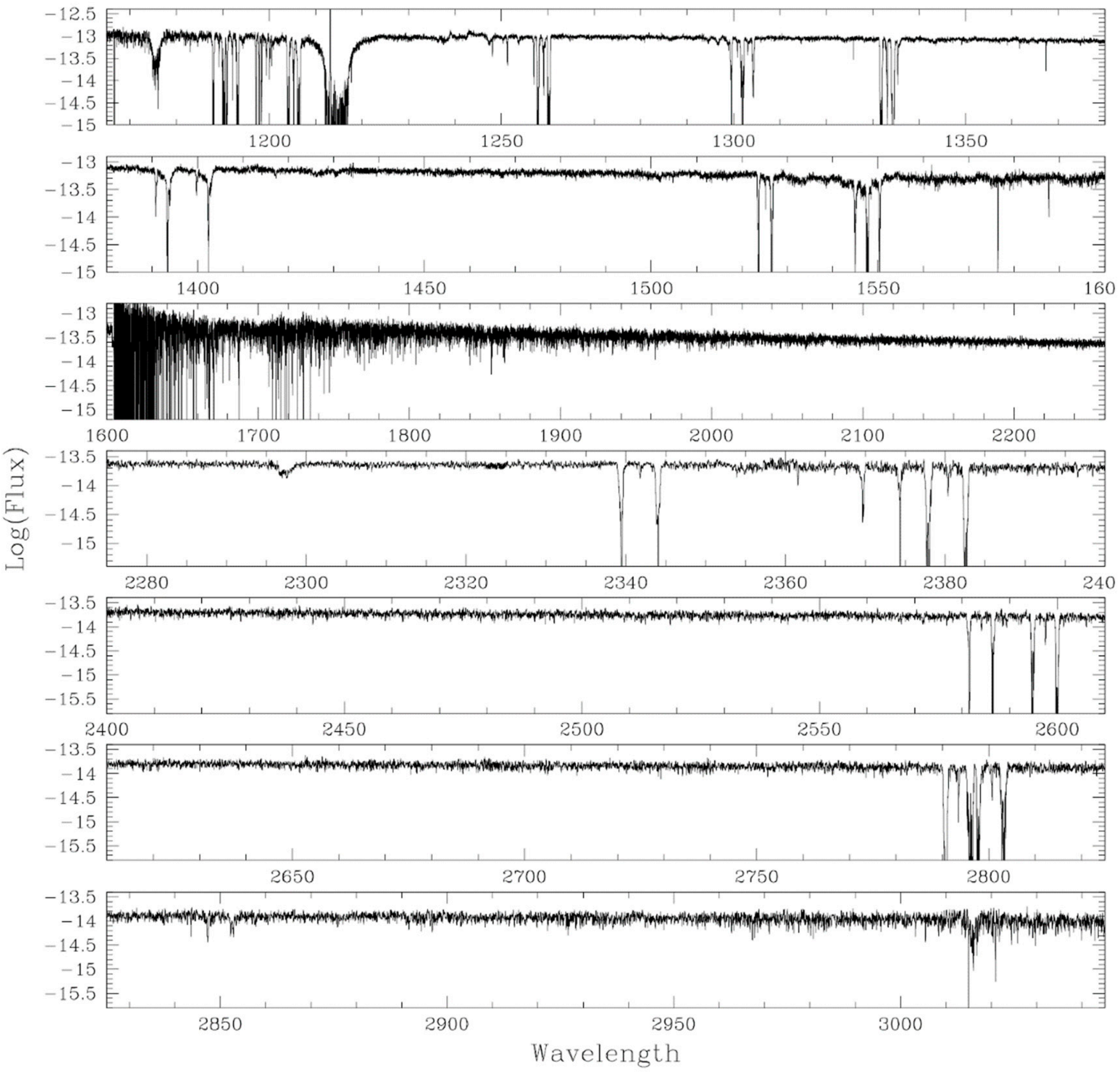

**Figure 3.** Hubble Space Telescope/Space Telescope Imaging Spectrograph (*HST/STIS)* spectrum of the central super star cluster in NGC 1705. The spectra were taken with the echelle gratings E140M and E230M, which have resolving powers of 46,000 and 30,000, respectively. Compare the quality of the spectrum to that shown in Figure 2. From [6].

## 2. Massive Stars in Nearby Star Clusters in the Era of Hubble

Giant H II regions are the most powerful sites of massive star formation found in the Local Group of galaxies [9]. They contain a large population of hot stars of spectral types O and B, and a substantial fraction of the known Wolf-Rayet (W-R) star population. Due to the significant numbers of hot, young, massive stars, which are evidence of a recent episode of star formation within the region, they serve as the analogs for more distant starbursts out to cosmological distances [10].

The 30 Doradus nebula in the Large Magellanic Cloud has often been called the "Rosetta Stone" of giant extragalactic H II regions [11]. Owing to its proximity, its contents of individual stars has been studied extensively (e.g., [12]) and compared to its global properties (e.g., [13]). Indeed, these and other studies have provided invaluable insight into the workings of high-mass star formation and the relation to starburst galaxies. The 30 Doradus nebula is a unique laboratory of star formation: approximately 25% of the entire massive star formation in the LMC occurs within a 15′ radius around 30 Doradus [14]. The central portion of the nebula is shaped by the cluster NGC 2070, which itself is subdivided into numerous smaller clusters of massive OB and W-R stars, the most luminous being the R136 cluster. The superb spatial resolution of the *HST*'s *STIS* permits a stellar census of R136 by performing low resolution far-UV *STIS* spectroscopy of R136 using adjacent long-slits for a complete coverage of the central 0.85 pc (see Figure 4). In [15], spectral types of tens of sources in the inner region were determined. They measured outflow velocities for 52 OB stars via C IV 1548-51, including 16 very early O2-3 stars. A complete Hertzsprung-Russell diagram for the most massive stars was generated, which suggests a cluster age of about 1.5 Myr. The stars in R136 were formed during the most recent star-formation episode within NGC 2070. The NGC 2070 cluster itself contains stars of older age and shows an age spread of ~6 Myr [16]. The individual spectra were then co-added to obtain the integrated UV spectrum of R136. The prominent He II λ1640 emission line indicates very massive stars, with initial masses in excess of 100 $M_\odot$. The extraordinarily strong emission is interpreted as being due to an IMF well above the canonical upper limit of 100 $M_\odot$. While spectroscopy is the royal method of determining the properties of the most massive stars, ancillary panchromatic photometry complements studies of the detailed star-formation history, particularly for less massive stars e.g., [17].

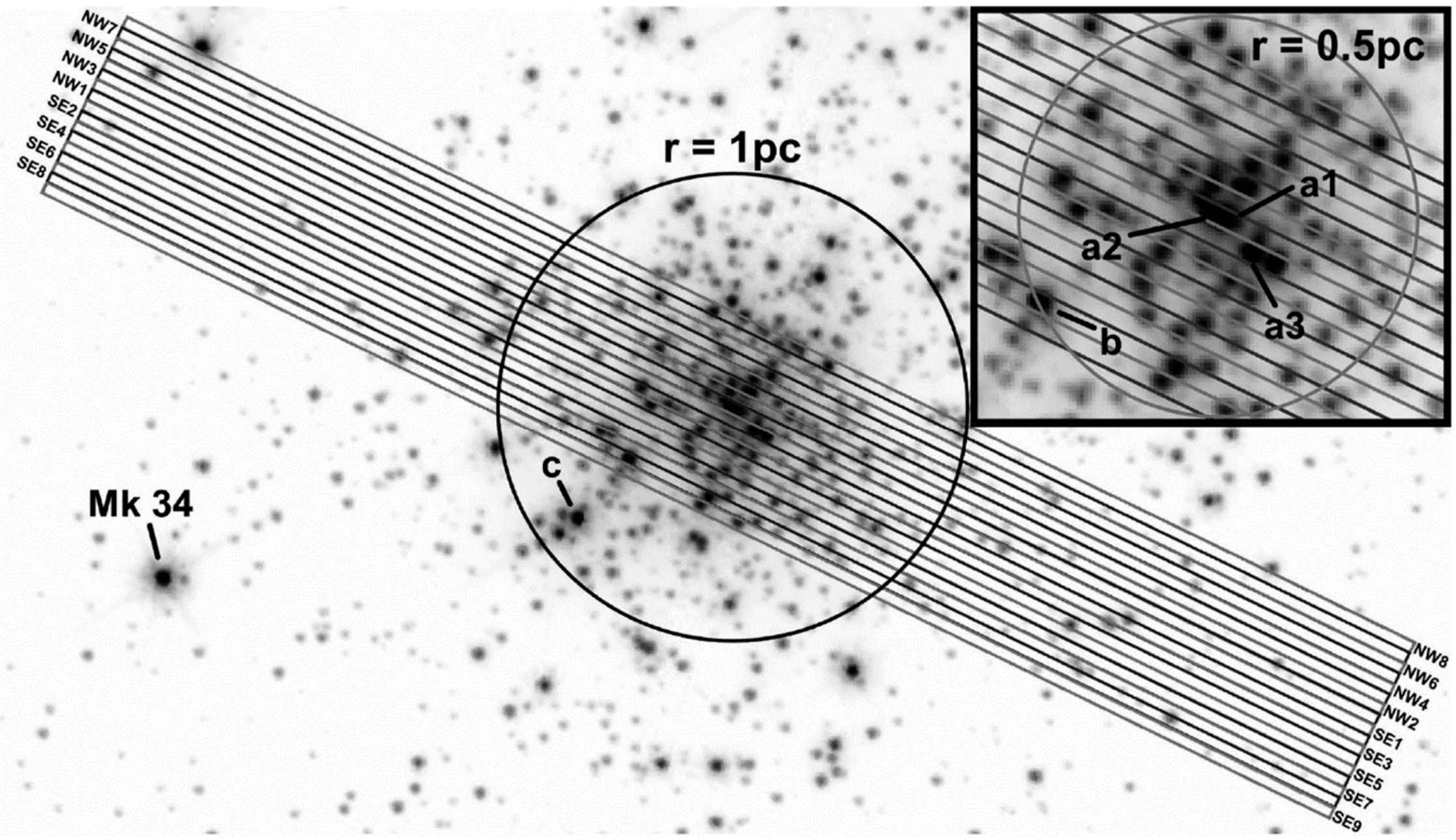

**Figure 4.** *HST/STIS* 52″ × 0.2″ apertures overlaid on an *HST* image of R136. The central region is shown in the insert at the top right, which has a radius of 2.05″, corresponding to 0.5 parsec at the distance of the LMC. From [15].

The 30 Doradus nebula is not fully representative of the giant H II region class. It is by far the densest, most concentrated H II region, setting it far apart from almost all other examples in the Local Group. (There are, however, giant H II regions with comparable or even more extreme properties at larger distances.) For instance, its luminous core R136 contains several thousand massive stars within 10 pc, which is about three times the number of stars that is distributed within the central 100 pc of NGC 595, the second most luminous H II region in M33 [18]. *HST* imaging of NGC 604, the most luminous H II region in M33, reveals an equally extended association of ~35,000 OB early-type stars within a field of 100 pc [19]. Both clusters lack the strong central concentration observed in R136 and at the same time occupy larger areas. The observational benefit is obvious: the less crowded populations enable stellar photometry and spectroscopy in M33 almost like in the LMC (albeit at lower flux levels), while the global non-stellar properties are much more reliably determined at the distance of M33: the 30 Doradus region extends over ~100 pc, or 500″ at the LMC distance, which exceeds the size of the *HST*'s spectroscopic entrance apertures by an order of magnitude. More importantly, astrophysical considerations raise concerns about generalizing the results for 30 Doradus only and applying them to other massive star formation regions.

In [20], a *STIS* two-dimensional spectral image with the UV G140L grating of the central concentration of OB stars in NGC 604 was obtained. The 2″-wide aperture encompassed approximately a 25″ × 2″ area in NGC 604 (see Figure 5). The resulting UV spectral image has a spectral resolution of about 2 Å and covers the wavelength range of 1150–1730 Å. The single spectral image permitted extraction and analysis of the 40 most luminous OB stars. These stars are young, with typical ages of ~3 Myr, which is somewhat older than the stellar ages in R136. Furthermore, NGC 604 host an older population of red supergiants whose age has been estimated to be ~12 Myr [21]. An atmospheric analysis of the hot, massive stars indicates a very high luminosity, and by implication a very high mass, of some of the stars [22]. A comparison of their location in the Hertzsprung-Russell diagram with theoretical evolutionary tracks suggests zero-age main-sequence masses in excess of 120 $M_\odot$ (see Figure 6). However, some of the stars may be unresolved binaries, which would lower their derived masses. If the stars are single, the slope of the derived IMF agrees with that of the classical Salpeter IMF. The UV light is dominated by very few UV-bright O stars. This suggests that incomplete sampling of the IMF could affect the interpretation of the UV luminosity if individual stars are no longer resolved in integrated spectra [23].

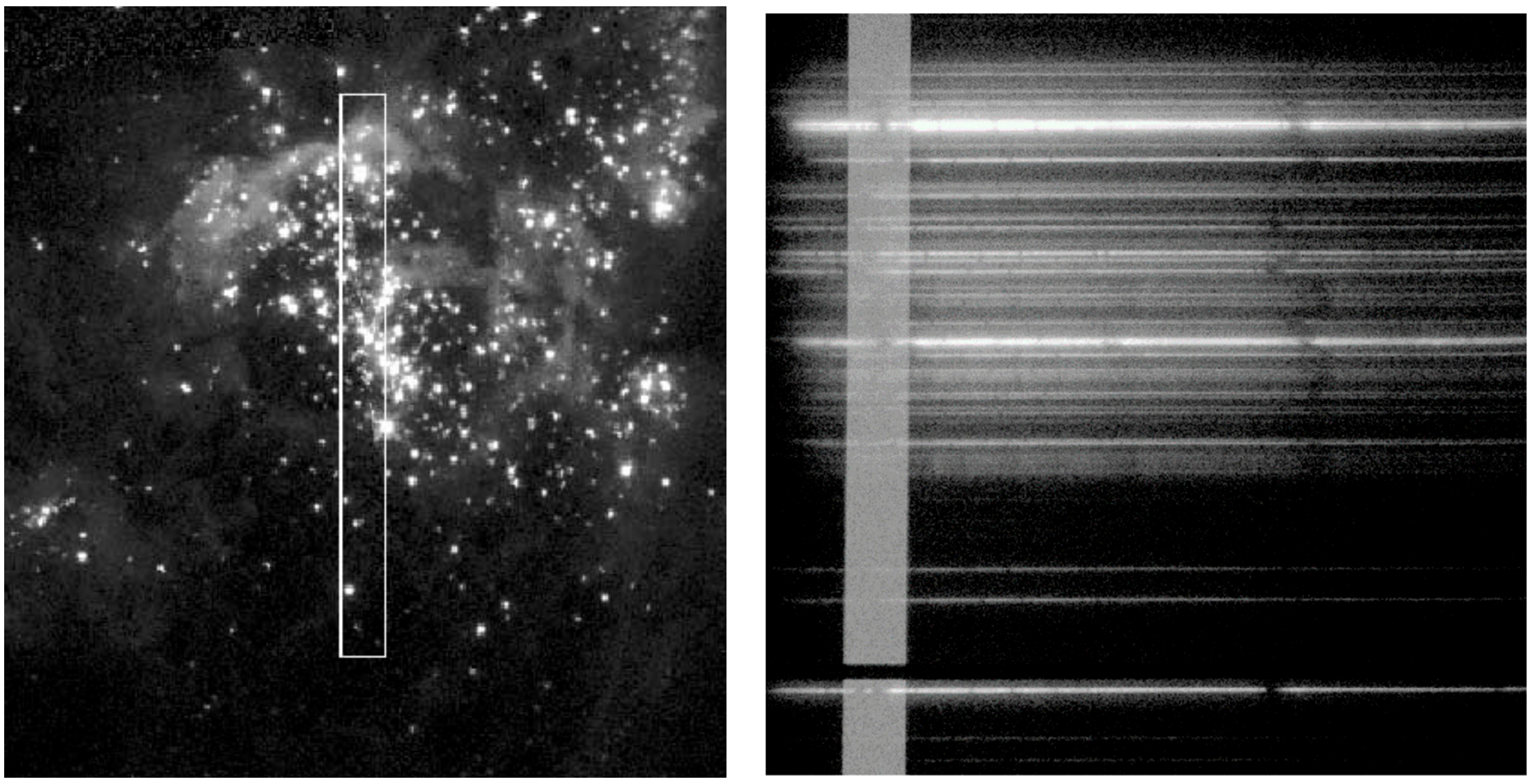

**Figure 5.** Left: the 2″- (corresponding to 8 pc) wide aperture of *STIS* superposed on a U image of NGC 604. Right: the *STIS* spectral image of OB stars in NGC 604, showing the individually resolved spectra. The vertical stripe to the left is geocoronal Ly-α. From [20].

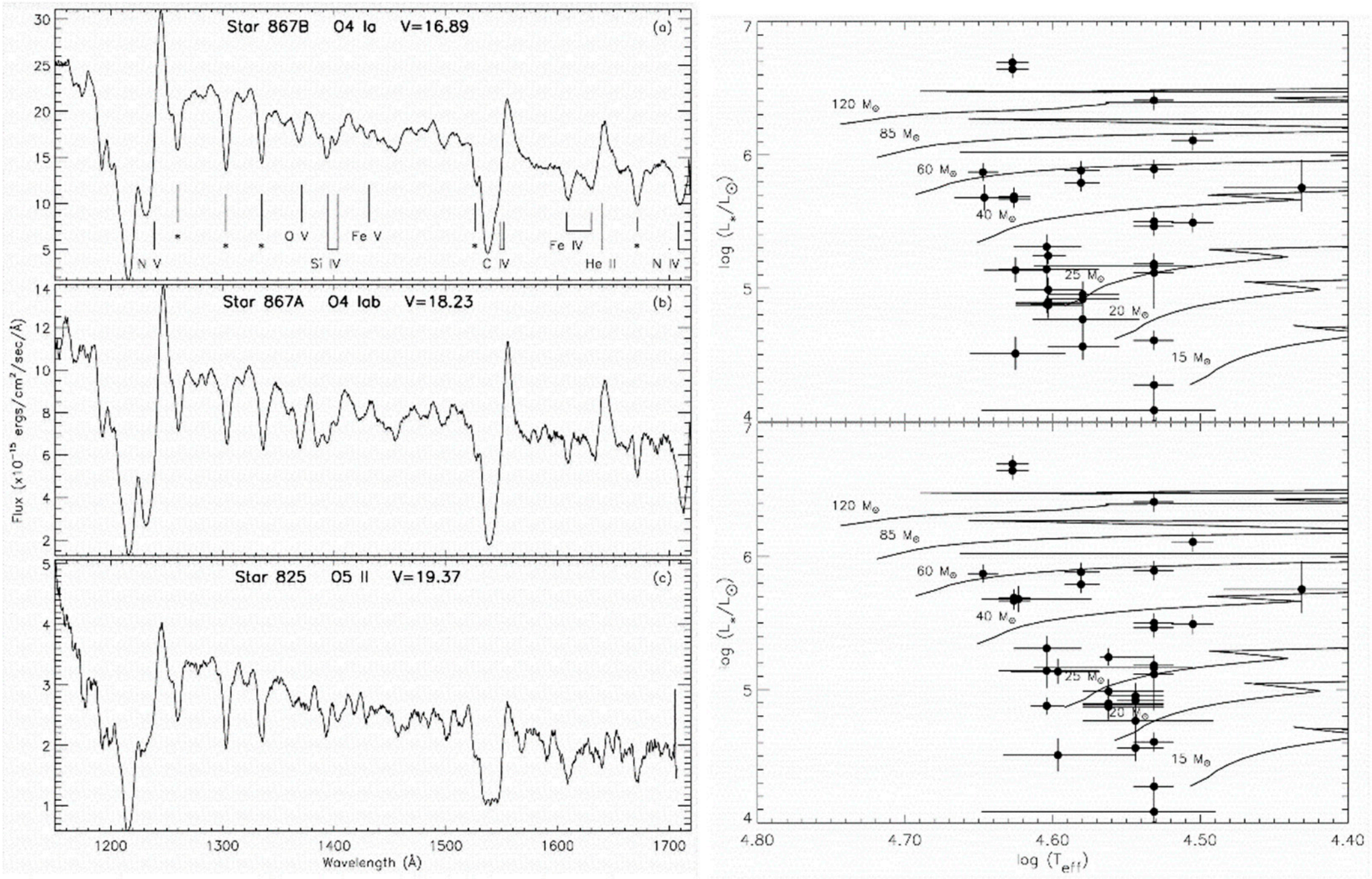

**Figure 6.** Left: examples of *STIS*/G140L UV spectra of the most luminous stars in NGC 604. Right: a Hertzsprung-Russell diagram of the most massive stars in NGC 604. The upper and lower figure are for two different calibrations of spectral type versus stellar parameters. From [22].

Spectroscopic studies in the UV of the resolved stellar content of young star clusters are largely restricted to our own Galaxy and the closest Local Group galaxies. This is simply dictated by the UV brightness of even the most luminous early-type stars and the limited spatial resolution of existing UV spectrographs. The unresolved stellar content of the more distant star clusters can be investigated from the integrated UV light.

## 3. Massive Stars in Integrated Star Cluster Spectra

Nearby star-forming galaxies make excellent training sets for investigating the starburst mechanism because their proximity affords superior spatial resolution. Their UV emission is typically dominated by a few massive star clusters and associated H II regions. Local star-forming galaxies have been shown to be close analogs to Lyman-break galaxies at high redshift in star-formation rate, UV colors, and spectral morphology [10]. Nevertheless, one should keep in mind that galaxies observed at high redshift have higher masses and different star-formation histories, and direct comparisons should be taken with a grain of salt.

The nearby starburst galaxy M83 (NGC 5236; $D$ = 3.7 Mpc) is a prototypical star-forming galaxy, whose characterizing property is its complex optical morphology. The central starburst in M83 extends over 20″ (360 pc at 3.7 Mpc) and displays a complex morphology. The optically detected starburst in M83 is confined to a semi-circular annulus between 3″ and 7″ (54 pc and 126 pc, respectively) from M83's optical center (see Figure 7). At *HST* resolution, a series of very young star clusters can be identified in the arc, whose ages range from a few to several tens of Myr. Star formation has not been coeval between the clusters but there is no directional age gradient, as suggested by an analysis of age-sensitive spectral lines in the UV spectra [24].

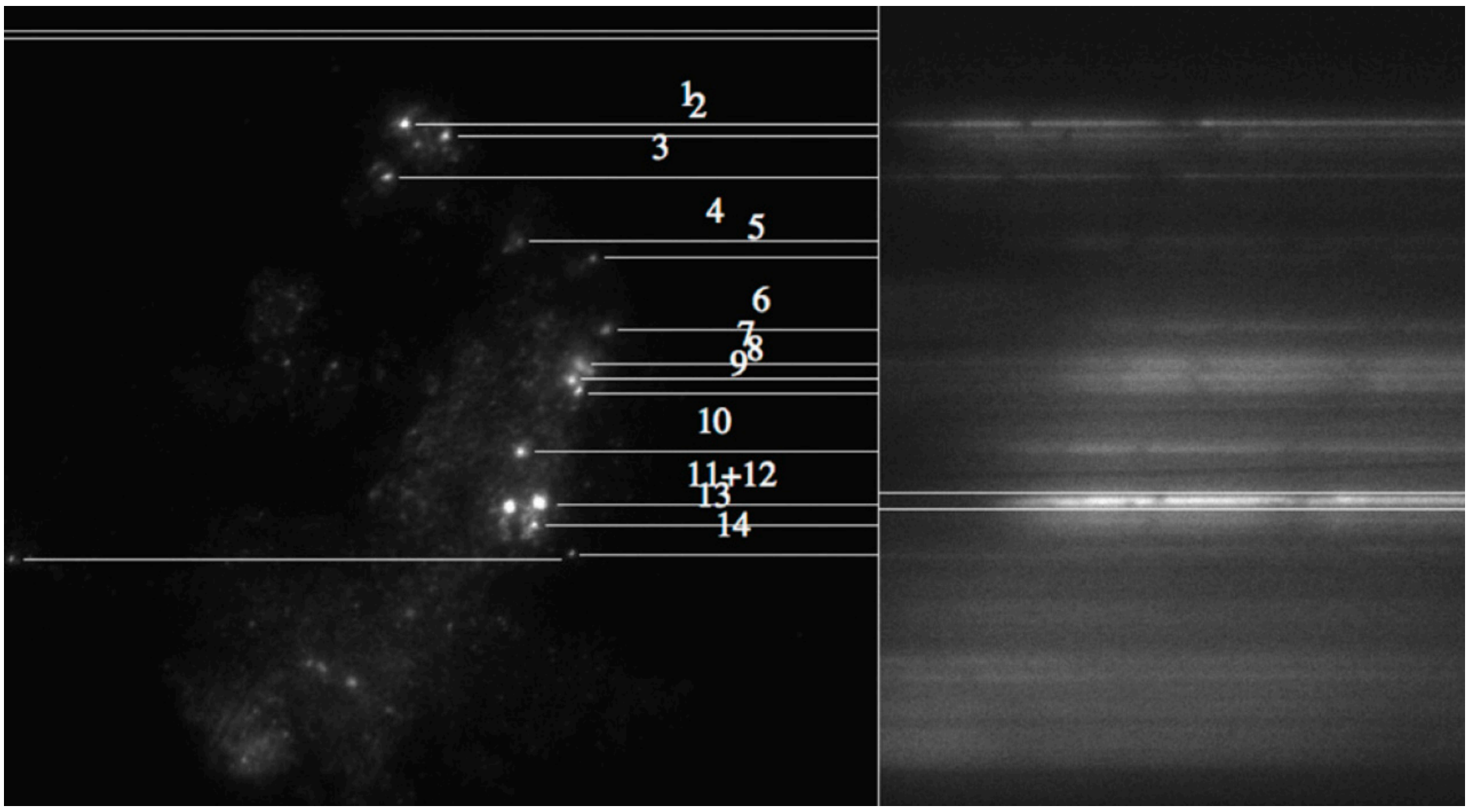

**Figure 7.** Left: an *STIS* UV image of the star-forming region near the center of M83. Many UV-bright star *clusters* can be seen. Compare with Figure 5, which shows individual *stars*. The field size is approximately 280 pc × 280 pc. Right: the corresponding spectra of the clusters obtained by using slitless spectroscopy in the region with *STIS*. From [24].

The space-UV lines in star-forming galaxies essentially come in four flavors: stellar-wind features, photospheric absorption lines, interstellar absorption lines, and nebular emission lines. The most prominent lines are usually the N V 1240, Si IV 1400, and C IV 1550 stellar-wind lines, which form in the expanding atmospheres of OB stars. These lines are broad (~2000 km $s^{-1}$), have blueshifts (~1000 km $s^{-1}$), and sometimes display P-Cygni profiles. Owing to the low wind densities (compared to photospheric densities), these lines are resonant transitions, i.e., their lower excitation energy is 0 eV. Because these lines are formed in a wind driven by radiation, which must overcome the gravitational well of the mass-losing star, the strength and shapes of the profiles depend on mass and are therefore sensitive to the IMF and star formation history. In addition, the line profile is also sensitive to the chemical composition since both stellar evolution and mass loss are strongly metallicity dependent [25].

Stellar photospheric absorption lines are also observed between 1200 and 3000 Å, where O- and B-star atmospheres are heavily blanketed, mostly by highly ionized iron and nickel lines [26]. Among the plethora of photospheric lines, there are several identified features (e.g., O IV 1342, Fe V 1363, Si III 1417, C III 1426/28, S V λ1502), many of which are blends of multiple lines. These blends are usually rather weak in comparison with the wind lines. Because their equivalent widths (EW) measure only a few Å (as opposed to 5–15 Å for the wind lines), only high signal-to-noise (S/N), moderate resolution spectroscopy can detect them. Yet, the photospheric features are invaluable because they provide an independent tracer of OB stars. In [27], several weak photospheric blends for metallicity determinations were calibrated. Their calibration was used by [28] to derive the stellar chemical composition of star-forming galaxies at redshift $z \approx 2$. Extrapolating a locally derived calibration to the high-redshift universe must be taken with care as such calibrations may change over cosmic time. While there is no evidence for evolution from $z = 0$ to $z \approx 2$, implications for galaxies at even higher redshift are unclear [29].

The ISM leaves a detectable signature in the UV spectra of star clusters as well. Strong interstellar absorption lines are formed by the ground-level transitions of the abundant atomic and ionic species of H I, C II, C IV, N I, N V, O I, Al II, Al III, Si II, Si III, Si IV, Mg I, Mg II, and Fe II. Most of these lines are optically thick, and therefore their strength is largely the result of the velocity dispersion of the gas in clusters and the covering factor of the ISM. These absorption lines are a unique tracer of the kinematics of the gas over a wide range of ionization energy.

Nebular emission lines can also be detected in the UV. However, in comparison with their counterparts at optical wavelengths, their EWs are usually small in all but the most metal-poor galaxies. The strongest lines are C III] 1907/09 and Si III] 1883/92. The C III] line has been used in combination with some of the weaker lines, such as O III] 1661/66, to derive the C/O abundance ratio [30].

In [31], in order to generate an atlas of all the scientifically useful UV spectra of young extragalactic star clusters and star-forming galaxies in the nearby universe, which were obtained with the Faint Object Spectrograph (*FOS*) and the Goddard High Resolution Spectrograph (*GHRS*) onboard the *HST*, the *HST* Archive was mined. In Figure 8, we reproduce their composite spectrum of the entire sample. This spectrum displays many of the spectral lines discussed in the prior paragraphs.

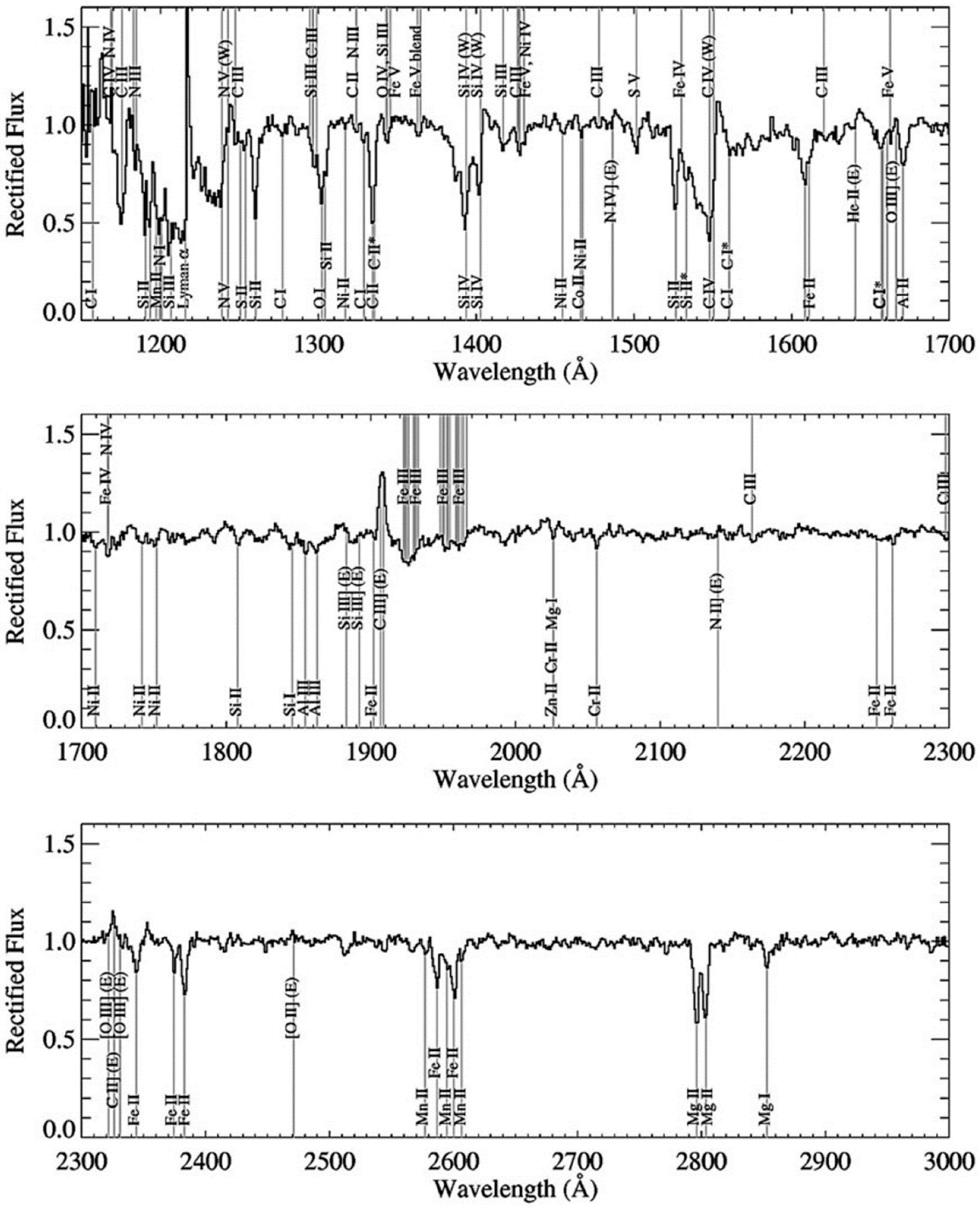

**Figure 8.** Composite spectrum constructed from a sample of star-cluster spectra obtained with the Faint Object Spectrograph (*FOS*) and the Goddard High Resolution Spectrograph (*GHRS*) onboard *HST*. Strong interstellar lines are identified below the spectrum; stellar-wind and photospheric lines are labeled above. The wind lines are marked with a "W". Nebular emission lines are identified with an "E". From [31].

The *Cosmic Origins Spectrograph* (*COS*) is the *HST*'s latest generation UV spectrograph. In [32], the *COS* UV spectra of young star clusters in nine nearby star-forming galaxies were collected. The most relevant global properties of the nine galaxies are summarized in Table 1 of [32]. Two different pointings were obtained in two galaxies. The galaxy sample covers an extensive range in oxygen abundance, galaxy type, and star-formation activity, providing a unique opportunity to document the chemical composition dependence of stellar and nebular properties at low redshift. The *COS* observations have an average spectral resolution of ~25 km $s^{-1}$, have an S/N ranging from ~10 to 30, and include the wavelength region ~1150–1450 Å. UV images of the nine galaxies, together with the position of the *COS* entrance aperture, are reproduced in Figure 9.

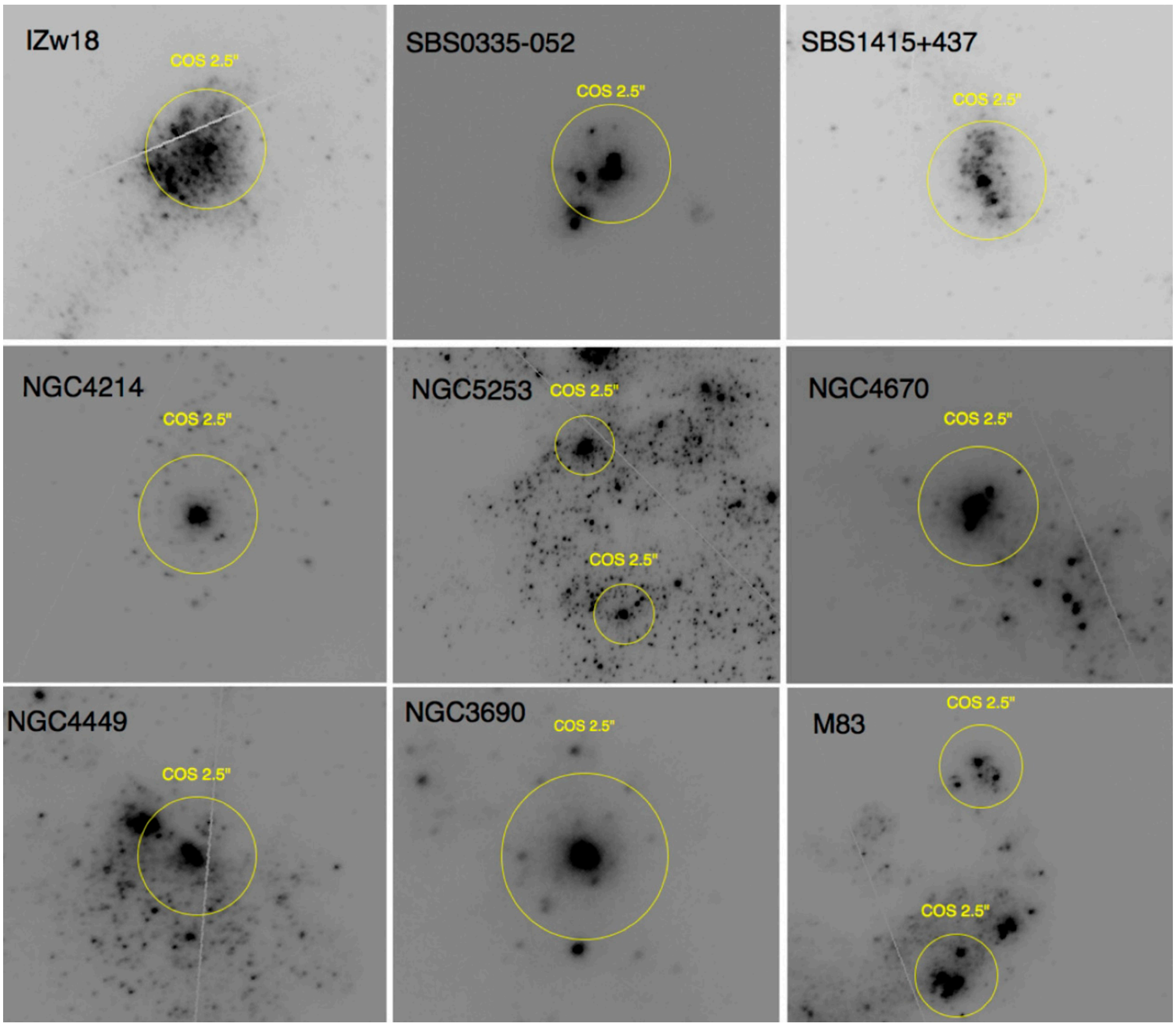

**Figure 9.** *HST* UV images of the star-forming galaxies hosting many massive young clusters. The overlaid circles indicate the 2.5″ aperture of the *HST Cosmic Origins Spectrograph* (*COS*). The entrance aperture encompasses a single (NG 4214) or multiple clusters (e.g., I Zw 18). The linear diameter of the field enclosed by the aperture is (from left to right, top to bottom): 220 pc, 650 pc, 165 pc, 37 pc, 46 pc, 280 pc, 46 pc, 587 pc, 58 pc. From [32].

Among the nine galaxies, I Zw 18 is the object famous for being the most metal-poor star-forming galaxy known in the local universe (12 + log(O/H) = 7.11 [33]). The system has two distinct morphological components that share the same H I envelope, the so-called primary and secondary bodies. The primary component itself is dominated by two star-forming regions, with the brighter region targeted by [32]. The *COS* spectra of I Zw 18 and the other galaxies are shown in Figure 9 and are plotted in Figure 10. The major stellar and interstellar spectral lines discussed before are clearly present in the spectra. Note the progression in line strength from the top to the bottom spectra, which represent a sequence of increasing chemical composition.

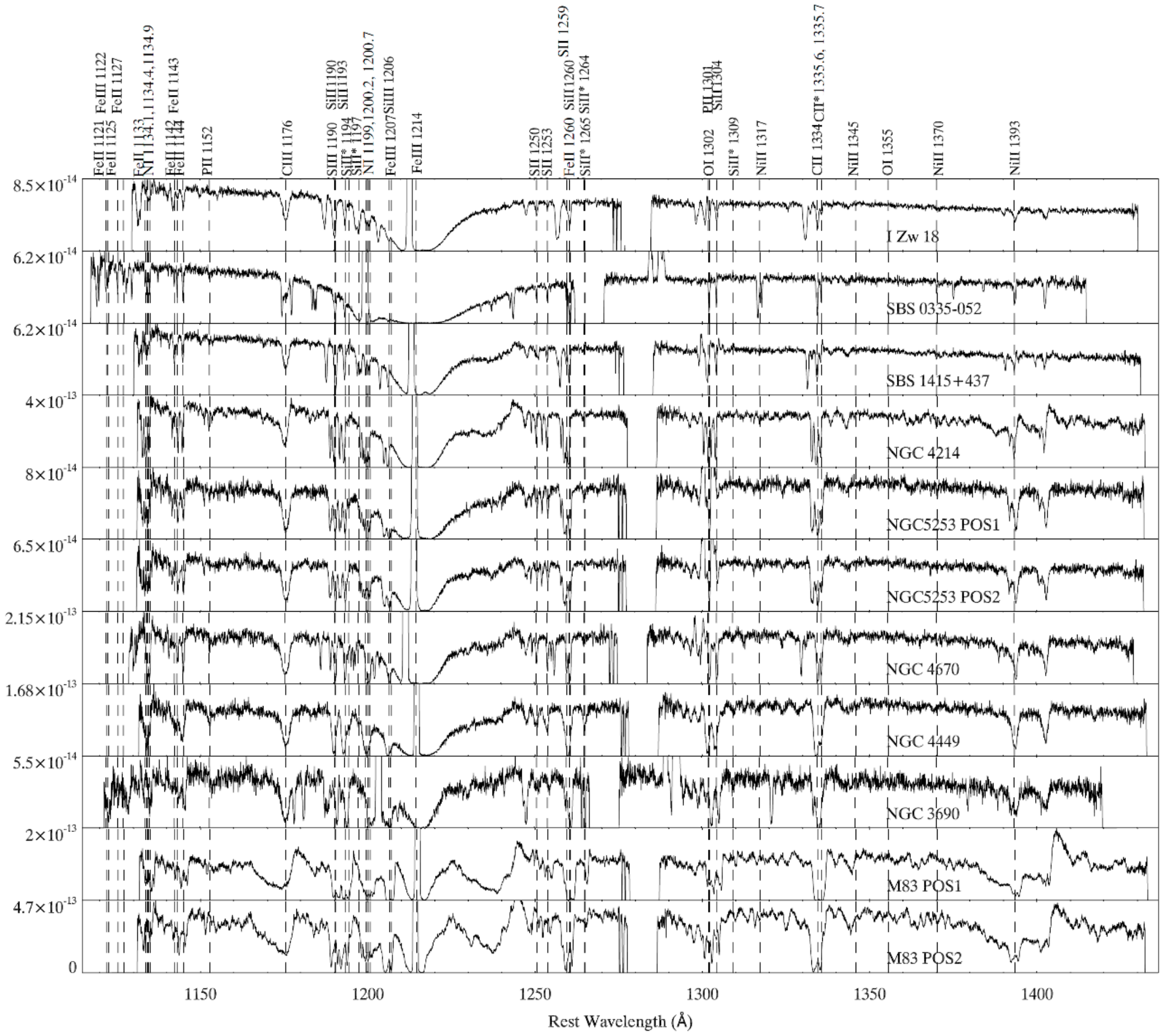

**Figure 10.** *COS* G130M spectra of the star clusters seen in the images in Figure 9. The spectra are arranged in order of increasing oxygen abundance from top to bottom. The major absorption lines in each galaxy are labeled in the uppermost panel. From [32].

Together with I Zw 18, SBS 0335−052 has been confirmed as one of the most metal-deficient star-forming galaxies in the local universe. The combination of the low oxygen abundance of SBS 0335−052 (~1/25 $Z_{\odot}$) and no indication of an underlying older population from deep imaging suggests that SBS 0335-052 may be an almost primordial galaxy in one of its first star-formation episodes [34]. However, deeper imaging is required for confirmation, as the detection of an old population would be close to the limits of current instrumentation.

The next galaxy in Figure 10, SBS 1415 + 437, is a metal-deficient blue compact dwarf (BCD) galaxy. Its reported oxygen abundance is ~1/12 $Z_{\odot}$. The galaxy was once thought to have experienced its first burst of star formation only 100 Myr ago [35]. Therefore, SBS 1415 + 437 was once considered as a prime example of a primeval galaxy candidate in the local universe. However, deep *HST* imaging revealed a previous population of much older stars, contradicting earlier claims of this galaxy being almost primordial [36].

NGC 4214 is a nearby (~3 Mpc) dwarf-irregular galaxy with two major star-forming regions containing hundreds of early-type stars as well as one very luminous star cluster. It has an oxygen abundance of 1/3 $Z_{\odot}$, similar to NGC 5253 and NGC 4670, which are also shown in Figure 10. Morphologically, NGC 4214 hosts multiple regions of active star formation, with luminous star clusters and large cavities cleared by stellar winds. The larger chemical composition and higher mass loss rates in this galaxy compared to the previous ones are indicated by the formation of a P Cygni profile in Si IV (see Figure 10).

NGC 5253 is located at a distance of ~3.8 Mpc. This starburst galaxy is known in the literature for being the first case of a nitrogen enriched H II region [37]. The presence of W-R stars is well established and thought to be the source of the nitrogen enrichment [38]. The two pointings obtained for this galaxy show quite similar spectra in Figure 10. Both spectra suggest the presence of luminous OB and W-R stars with strong stellar winds.

The next galaxy, NGC 4670, is an amorphous, oxygen-deficient ($Z \approx 1/3\ Z_\odot$), BCD galaxy. The UV spectrum is similar to that of NGC 4214, suggesting similar massive star content with vigorous star formation and an underlying older population [2].

NGC 4449 is a Magellanic-type irregular galaxy. Star formation occurs at elevated levels across the entire galaxy at about twice the rate of that of the LMC. H I studies of NGC 4449 indicate a rather complex structure, with extended gas streams around the galaxy. The H II region oxygen abundance is ~1/2 $Z_\odot$.

NGC 3690 has close to solar oxygen abundance. This galaxy is an example of an extremely disturbed system that is interacting or merging with the close-by galaxy IC 694 [39]. NGC 3690 has the highest redshift of the galaxies in Figure 10 ($z \approx 0.01$). Therefore, the Milky Way foreground absorption lines are well separated from the absorption lines intrinsic to the galaxy. For instance, the absorption at ~1322 Å is the blueshifted C II 1335 Milky Way absorption line.

M83 (see also Figure 7) is a grand-design spiral with Hubble type SAB(s)c and a central starburst [40]. The two *COS* spectra were taken at this central region, for two UV sources separated by about 150 pc (see Figure 9). M83 is the most-metal-rich galaxy among the sample, with a super-solar oxygen abundance. The two spectra at the bottom indicate strong stellar winds via the P Cygni profiles of C III 1175, N V 1240, and Si IV 1400.

The integrated spectra of young massive star clusters and star forming galaxies hold the key to revealing the content of massive stars. A comprehensive analysis of the far-UV spectra of 61 star-forming galaxies has been performed by [41]. A major tool for determining the relevant properties is population synthesis, which is the topic of the next section.

## 4. Population Synthesis Models for Massive Star Populations

The fundamental principle of population synthesis [42] is relatively straightforward. Stars are formed according to an assumed IMF and with a certain time dependence. Stellar evolution models give the relation between mass and luminosity, where the latter is the observed parameter, and prescribe how this relation evolves with time. Spectral libraries and other ingredients are then associated with each star in the Hertzsprung-Russell diagram in order to provide a broad range of predicted properties. Other ingredients can be the parameters of the ISM, such as dust attenuation, the morphology of the star-forming region, or corrections for stochastic effects. The final product is, among others, a complete spectral energy distribution (SED) of an individual star cluster or of an entire galaxy. Figure 11 from [43] is a schematic of these building blocks. The example in this figure uses a galaxy SED in the presence or absence of dust. Dust absorbs the stellar non-ionizing UV radiation and re-radiated it in the mid- and far-infrared (IR), producing a characteristic "double-hump" [44].

Widely used public model packages are BPASS [45], CIGALE [46], GALEV [47], GALEXEV [48], MILES [49], PÉGASE [50], and Starburst99 [51]. A comprehensive review of fitting simulated SEDs to observations has been published by [52]. In the following, I will briefly discuss the three major building blocks (star formation, stellar evolution, and stellar libraries) and address some open issues.

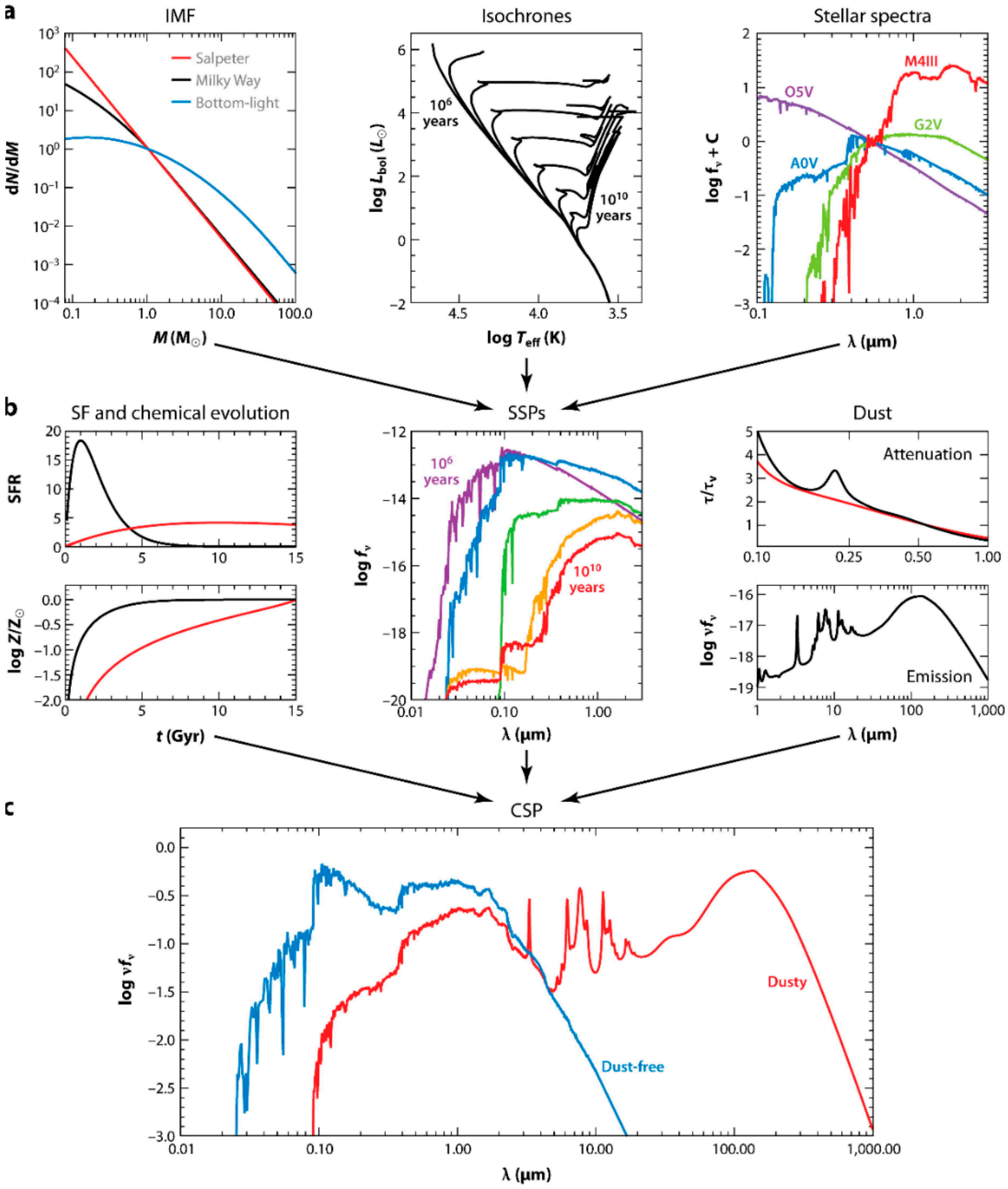

**Figure 11.** Schematic showing the principal components of population synthesis models. (**a**): input IMF, isochrones, and spectral library; (**b**): adopted star-formation history, single stellar population (SSP) calculated from the top row ingredients, and dust attenuation; (**c**): final spectrum for a continuously forming population (CSP) by combining the components in the middle row. From [43].

## *4.1. Star Formation and Related Properties*

Synthesis models for massive stars are age-degenerate if the star-formation episode lasts for more than about 10 Myr. This duration corresponds to the evolutionary time scale of O and W-R stars, which produce most of the UV photons. Stellar populations with constant, continuous star formation older than this will have indistinguishable UV properties because they are in an equilibrium between stellar birth and death. Ages in this regime are typically determined for the global populations in star-forming galaxies [53]. In [54], the evidence of IMF variations in massive-star populations was reviewed. The observations are consistent with a Salpeter-like IMF for stars at the high-mass end (see Figure 12). The previously stated caveat of possible stochastic or incomplete sampling applies to the interpretation of these data as well [23].

The chemical composition of the newly formed stars can affect the model predictions for the UV light in a complex way. In general, fewer metals will lead to a harder UV spectrum due to less line-blanketing. However, when two populations with solar and 1/7$^{th}$ solar chemical composition are compared in models using evolutionary tracks with and without rotation, there is little difference.

This results from two counteracting effects: less blanketing produces harder spectra, but, at the same time, metal-poor evolution single-star models with rotation [56] produce fewer very hot W-R, whose high-energy photon output is missing.

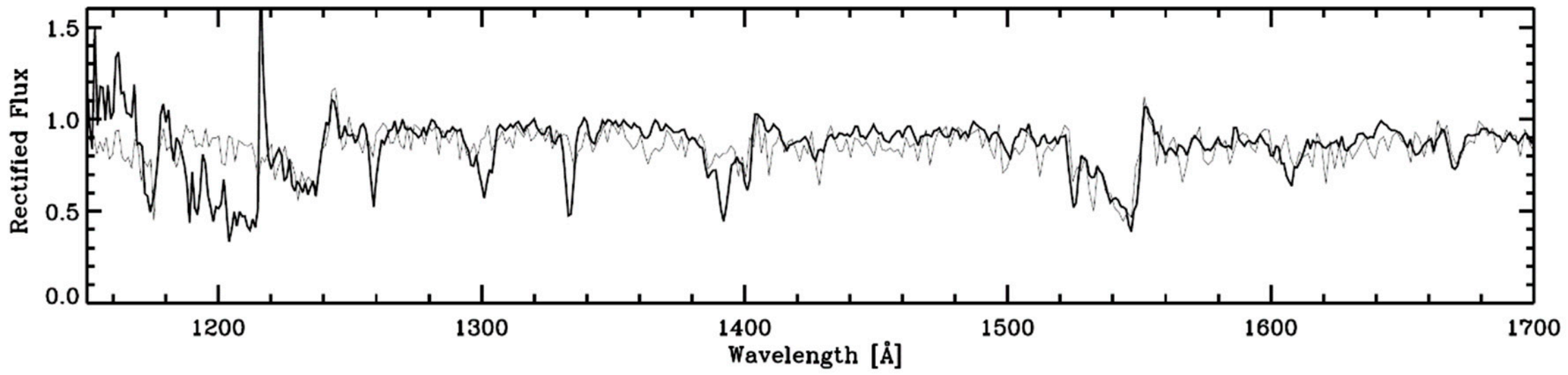

**Figure 12.** Comparison of the average spectrum of the 46 local star-forming regions shown in Figure 8 (dark spectrum) and a simulated spectrum following a Salpeter IMF with stars up to 100 $M_\odot$ (light spectrum). Models and data of the wind features of N V 1240, Si IV 1400, and C IV 1550 are in excellent agreement. The narrow absorption features (e.g., at 1260 Å, 1335 Å, and blended with the Si IV 1400 P Cygni profile) are interstellar and are not modeled in the synthetic spectrum. From [55].

### *4.2. Stellar Libraries*

UV spectral libraries can be based on observational or theoretical stellar spectra. The trade-offs of using either library have been discussed by [57]. Arguments favoring empirical libraries are:

- Laboratory data are sometimes incomplete or uncertain, and therefore the quality of the line lists may be insufficient for the computation of model atmospheres.
- The computational effort for producing a large model set can be challenging. This may limit the available parameter space.
- Departures from local thermal equilibrium (LTE) are important for hot stars. This can greatly increase the computational effort.
- Hot luminous stars have outflows and are extended. Therefore, the models must account for sphericity effects and possibly include hydrodynamics.
- Deviations from spherical symmetry can be relevant, in particular if the stars are not single.

On the other hand, the shortcomings of observational libraries are:

- Massive stars are rare and some stellar species (e.g., extremely metal-poor massive stars) are simply not found locally.
- Telescope time is precious, whereas computer time is comparatively inexpensive (and keeps becoming less expensive). Allocating telescope time for generating an extensive stellar library is often not considered high science return.
- Dust reddening for massive stars can be significant. UV data in particular often require large reddening corrections.
- An even more serious issue concerns interstellar absorption lines in the UV. Such lines often contaminate stellar resonance lines.
- An often-neglected issue is the need of a calibration of the spectral-types against temperature when the observed spectra are linked to evolution models. This relation is based on models, so that ultimately observational libraries are model dependent as well.

These issues affect different stars differently. The concerns about the reliability of theoretical libraries are mostly relevant to cool stars. The low temperatures of cool stars magnify the shortcomings of current atmosphere modeling so that late-type stars, including red supergiants, are better represented by empirical spectra in spectral evolution models. The reverse is true for hot stars. In this case, the models are rather mature [58] and any uncertainties are less of a concern when compared to the disadvantages of empirical libraries, such as limited parameter space and contamination by interstellar lines and dust reddening. Following these arguments, the Starburst99 synthesis code uses a fully theoretical UV stellar library as its default [59]. Figure 13 compares this theoretical library to an empirical library based on *IUE* spectra. The stellar lines are in excellent agreement but notice the narrow interstellar lines contaminating the *IUE* data. If these synthetic models were used for a study of the stellar content using automatic fitting routines, the presence of the interstellar lines would introduce a bias unless masked out.

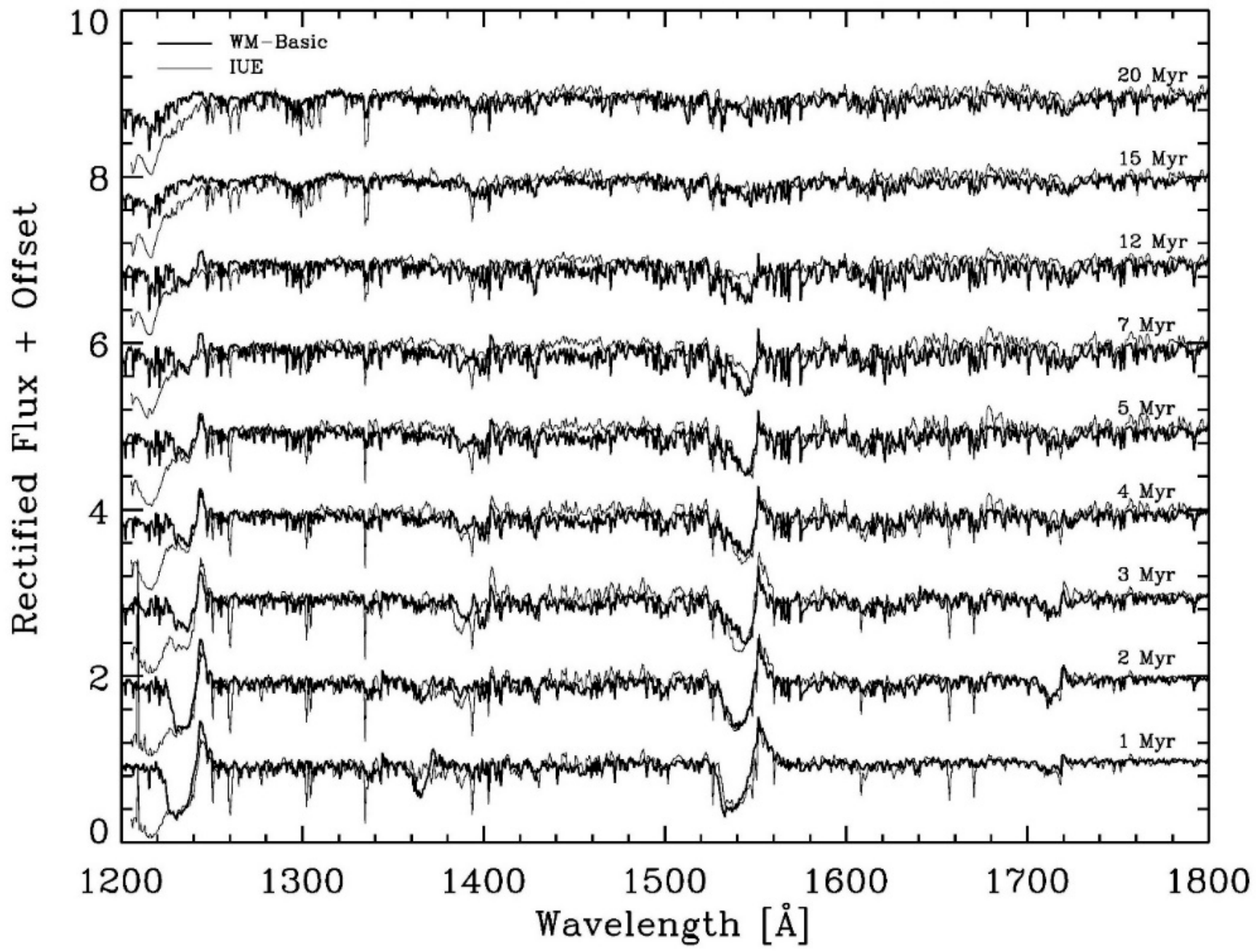

**Figure 13.** Comparison of the simulated spectra of a single stellar population with a solar chemical composition (thick lines) and an observational *IUE* library of Galactic stars (thin). The age steps from 1 to 20 Myr are labeled on the right. From [59].

## *4.3. Stellar Evolution*

The evolution of massive stars is still poorly understood. There is ample observational evidence that the available models are rather incomplete [60]. Massive stars show chemically processed materials, such as nitrogen from the CN cycle, early in their evolution. Previously, strong mass loss was invoked to remove the surface layers, but a downward revision of the empirical mass-loss rates led to the recognition that steady stellar winds are insufficient for the mass removal [61]. Alternatively, mixing processes could operate, which transport processed matter from the convective core to the outer layers. Differential rotation can induce such mixing. Evolution models for rotating massive stars have become widely available [62–65]. A comparison of models for rotating and non-rotating stars is presented in Figure 14. The tracks of stars with masses above ~20 $M_{\odot}$ are the most affected by rotation. Rotating stars have higher surface temperatures due to their lower surface opacity (hydrogen, which is the dominant opacity source, is reduced), and they are more luminous due to their enlarged convective core. The combination of the two effects leads to an ionizing luminosity of rotating stars that surpasses that of non-rotating stars by a factor of several [55]. The tracks shown in Figure 14 are for solar chemical composition; even more dramatic differences between non-rotating and rotating models arise in metal-poor stars. Since our knowledge of the rotation velocities of massive stars is still incomplete, the choice of the evolution models is the major uncertainty in synthesis models. Moreover, single-star models may not be applicable in all cases, and binary evolution models should be considered [45].

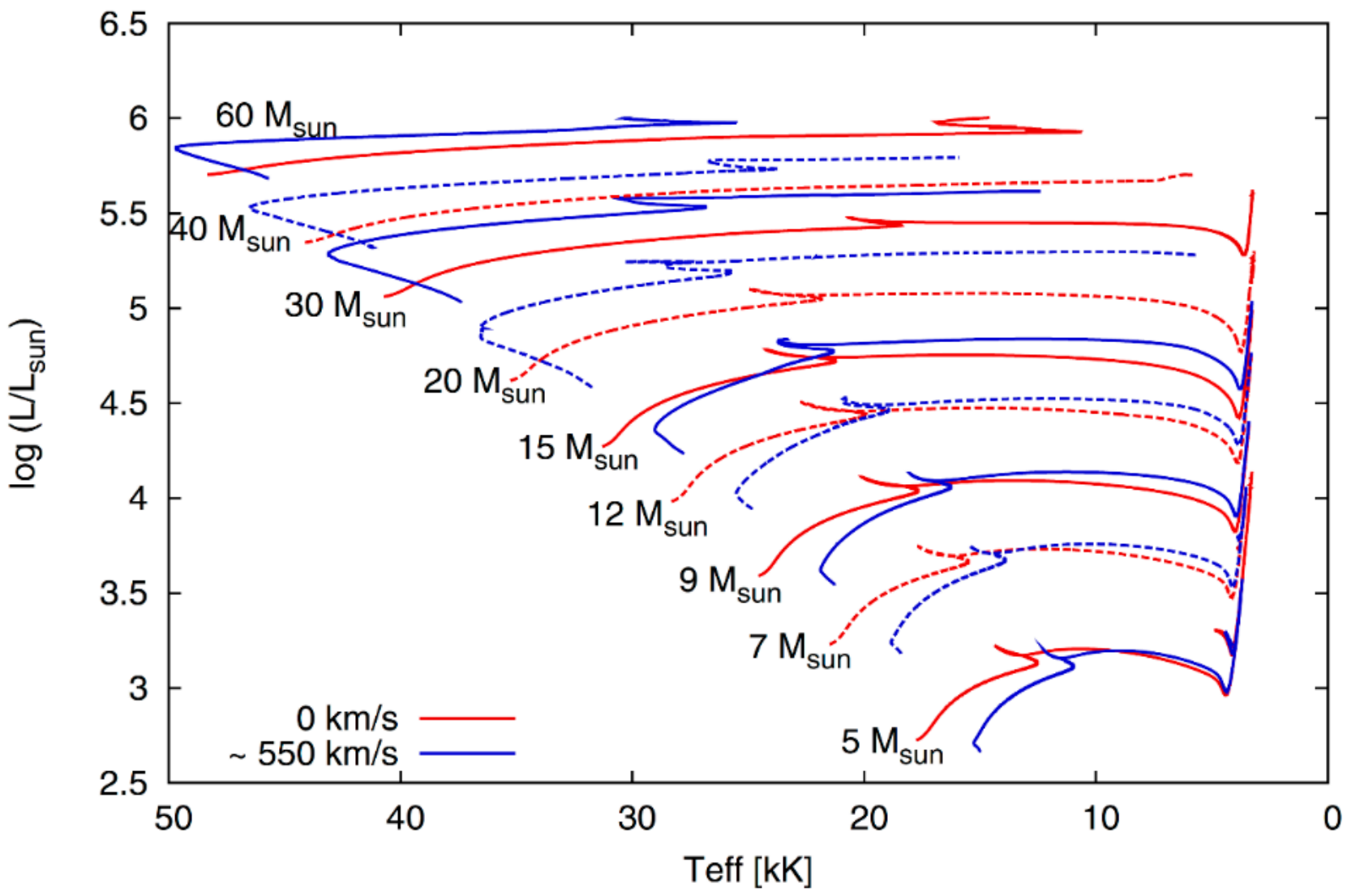

**Figure 14.** Comparison of evolution models at solar metallicity with rotation velocities on the zero-age main-sequence of 0 and 550 km $s^{-1}$. The velocities may be higher than actually observed in stars but were chosen to illustrate the effect of rotation on the evolutionary tracks. From [62].

## *4.4. Stellar Multiplicity*

Most massive stars are found (or are predicted) to have one or more companions. The observed fraction of stellar systems with two or more stars ranges from about 50% for low-mass stars to almost 100% for stars with masses above 10 $M_\odot$ [66]. The relevant aspect for population synthesis of unresolved stellar populations is the interaction history of the stellar components. Do the stars evolve as effectively single, and if they do not, how does interaction change the course of stellar evolution? Observations of individual stars in the Galaxy and the LMC can provide crucial guidance for the importance of different evolutionary channels [67]. Results from the VLT-Flames Tarantula Survey suggest that the 30 Doradus region has an intrinsic binary fraction of 51% [68].

After applying careful incompleteness corrections, [67] found that about 30% of all massive stars evolve as effectively single, whereas in 70%, binarity affects the evolution to some degree. After the stars have left the main sequence and their radius increases, the primary fills its Roche lobe and transfers material to the secondary. The mass-accreting secondary may spin up, and in the most extreme case the primary may get stripped of its envelope. For very close binaries, the two components may even merge. In [69], the evolution of close massive binaries after Roche-lobe overflow was modeled. Their models suggest dramatic changes in the extreme UV region of the spectrum. At ages older than 10 Myr, when single stars fade in the extreme UV, the stripped primary star has been transformed into a hot helium star with a high ionizing luminosity. If such stars are formed in a stellar population, the resulting ionizing luminosity surpasses that of single-star populations by large factors. Of course, the absolute values would be rather low for either population due to the advanced age. Analogous predictions are made by the BPASS models of [45].

## *4.5. Very Massive Stars*

Regions of massive star formation, such as star clusters and giant H II regions, are excellent training grounds for population synthesis models. A comparison of the observations of R136, the center of 30 Doradus in the LMC and NGC 5253, a nearby BCD galaxy, reveals shortcomings in the synthesis models [70]. R136 and the luminous star clusters in NGC 5253 have young ages of less than 2 Myr. Unexpectedly, broad He II 1640 emission is detected in their UV spectra (see Figure 15). This line (and its optical counterpart at 4686 Å) has traditionally been interpreted as due to evolved W-R stars. Stellar evolution models do not predict the formation of these stars at such an early age.

The failure of the synthesis models to match the data may be due to the combination of two effects: (i) Very massive stars with masses of 200 $M_\odot$ or higher can form stochastically even if they are not predicted by a power-law IMF for a cluster with relatively low mass. (ii) Very massive stars may be subject to strong internal mixing on the main sequence. These stars can evolve quasi-homogeneously and produce spectra mimicking those of classical W-R stars. Exploratory stellar evolution models for a limited stellar parameter range have been released [71] but they are not yet included in population synthesis models.

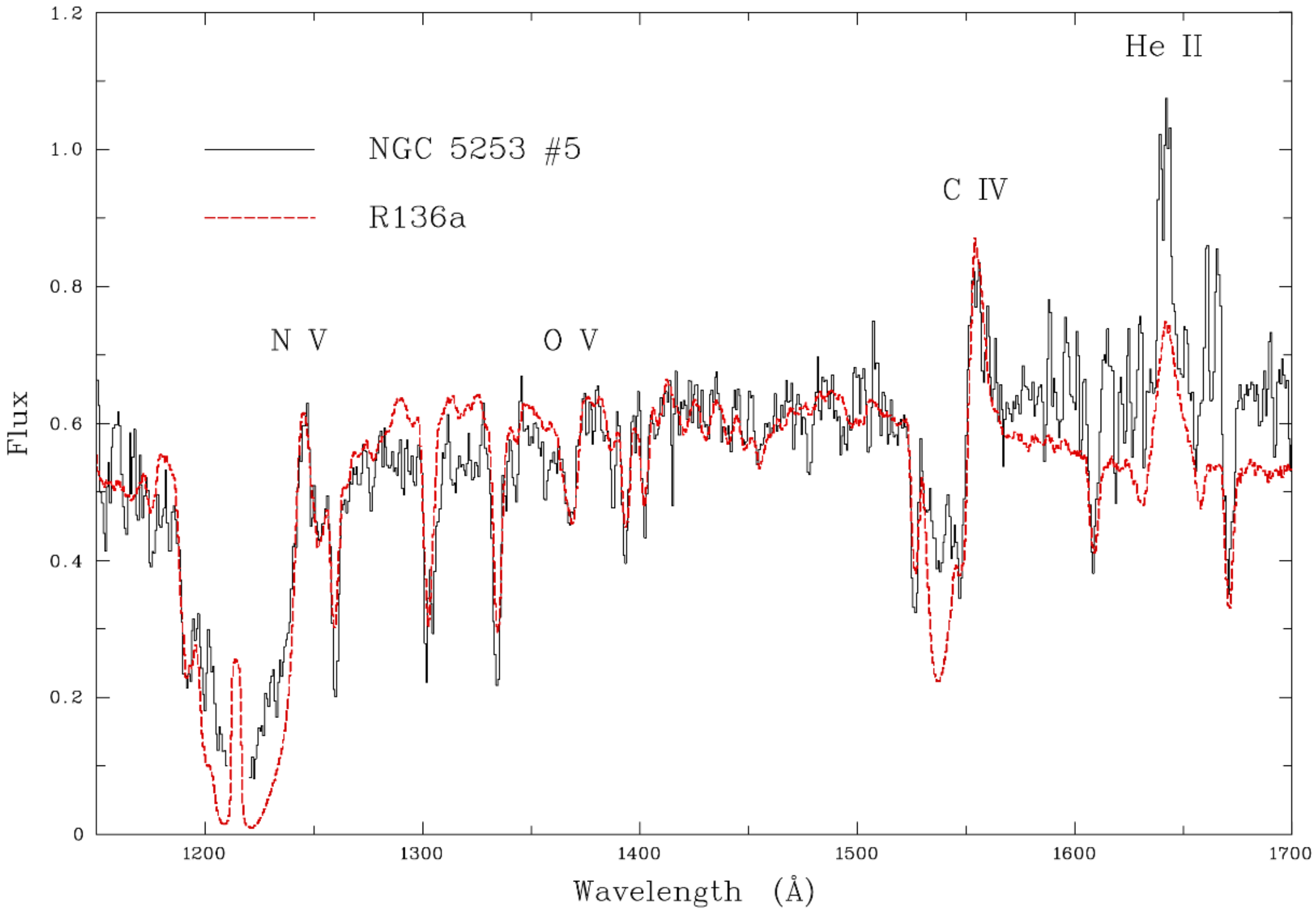

**Figure 15.** Comparison of the UV spectra of R136a, the central region of 30 Doradus, and the massive star cluster #5 in NGC 5253. Note the strong, broad He II 1640 line, which indicates He-enriched stars with dense stellar winds. From [70].

## 5. II Zw 40—A Case Study

In this section, I will focus on a case study of an individual object: a luminous star cluster and the associated giant H II region in the galaxy II Zw 40 (= UGCA 116). This galaxy, together with I Zw 18, is the original member of the class of BCD galaxies. Its oxygen abundance is close to that of the Small Magellanic Cloud (SMC). The class has initially been defined by [72], who demonstrated that its optical spectrum cannot be distinguished from that of extragalactic H II regions. BCDs and related dwarf starburst galaxies have since been classified as objects that are described by their blue optical colors, small sizes of less 1 kpc, and low luminosities of $M_B > -18$ [73]. II Zw 40 ($D \approx 11$ Mpc) is outstanding when discussed in the context of the general galaxy population in the local volume. II Zw 40 has the second highest H$\alpha$ equivalent width of 451 Å among the 436 galaxies in the 11 Mpc sample of [74,75].

The UV spectra of the H II region and its ionizing cluster (hereafter called SSC-N) were acquired with the *HST*'s *COS* using the G140L grating [76]. This instrument configuration covers the wavelength range 1150–2000 Å and provides a spectral resolution of approximately 0.5 Å. The position of the *COS* entrance aperture is shown in Figure 16. The stars providing the ionizing photons are concentrated in SSC-N, which is essentially unresolved in the UV (left part of the figure). In contrast, the optical H$\alpha$ image (right) suggests extended gaseous emission within the *COS* aperture, as well as diffuse emission covering hundreds of parsecs. The entire region is in the center of II Zw 40, whose extended (several kpc) tidal tails are relics of previous interaction and the merging of two dwarf galaxies [77].

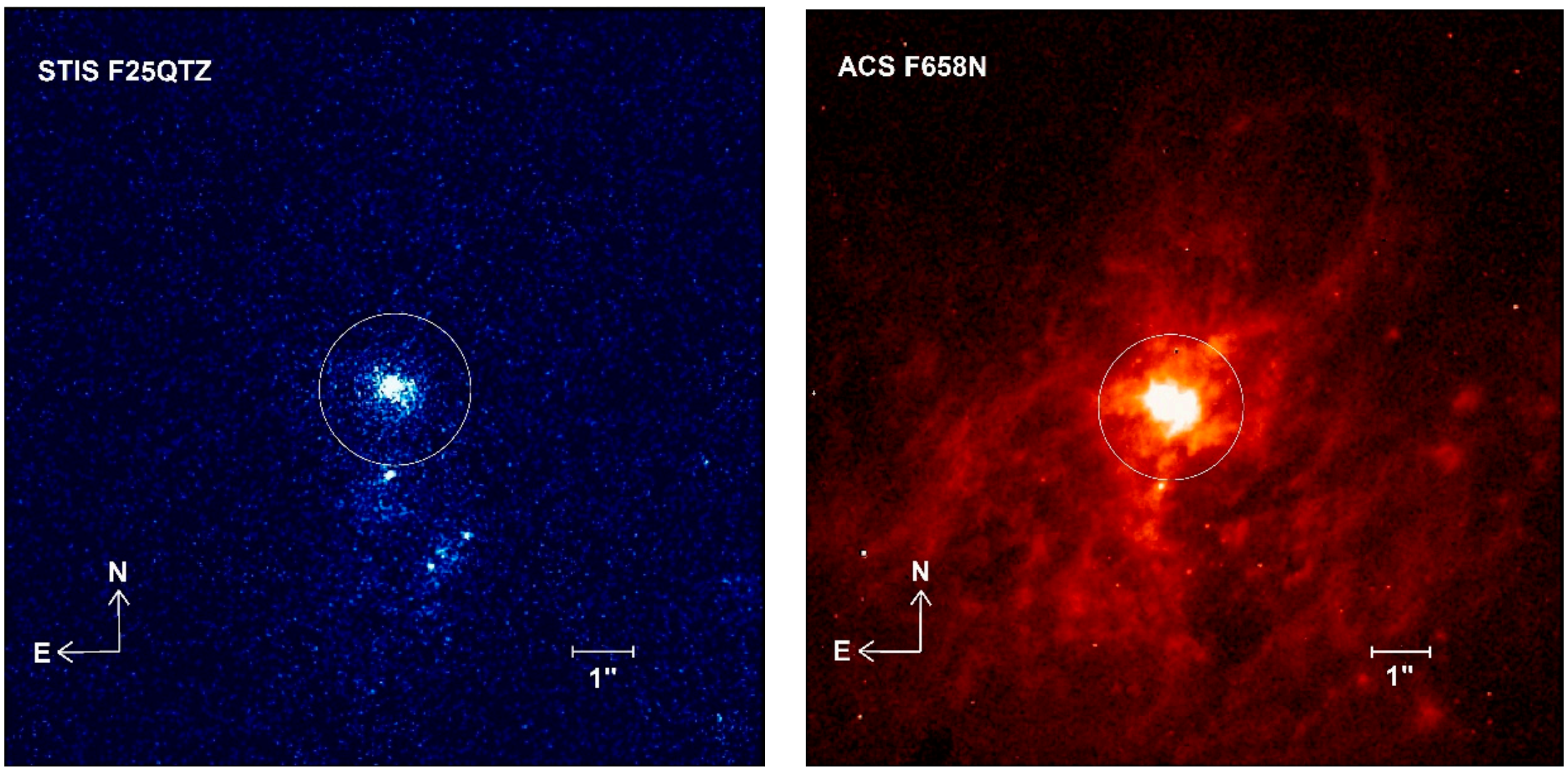

**Figure 16.** *STIS* UV (left) and Advanced Camera for Surveys (*ACS*) Hα (right) images of the star cluster and H II region in II Zw 40. Table 1600. Å. The circle denotes the position of the *COS* aperture, which encompasses a physical area of 135 pc in diameter. From [76].

The processed COS spectrum of SSC-N is reproduced in Figure 17. The spectrum has not been corrected for reddening, and the wavelength scale is in the observed frame. The significant Milky Way reddening of $E(B-V)_{\rm MW}$ = 0.73 is responsible for the decline of the flux at shorter wavelengths. The two most conspicuous emission lines are geocoronal Lyman-α and O I 1304. The width of the two lines is the result of them completely filling the COS entrance aperture. Any intrinsic Lyman-α emission from II Zw 40 SSC-N would be veiled by the geocoronal line.

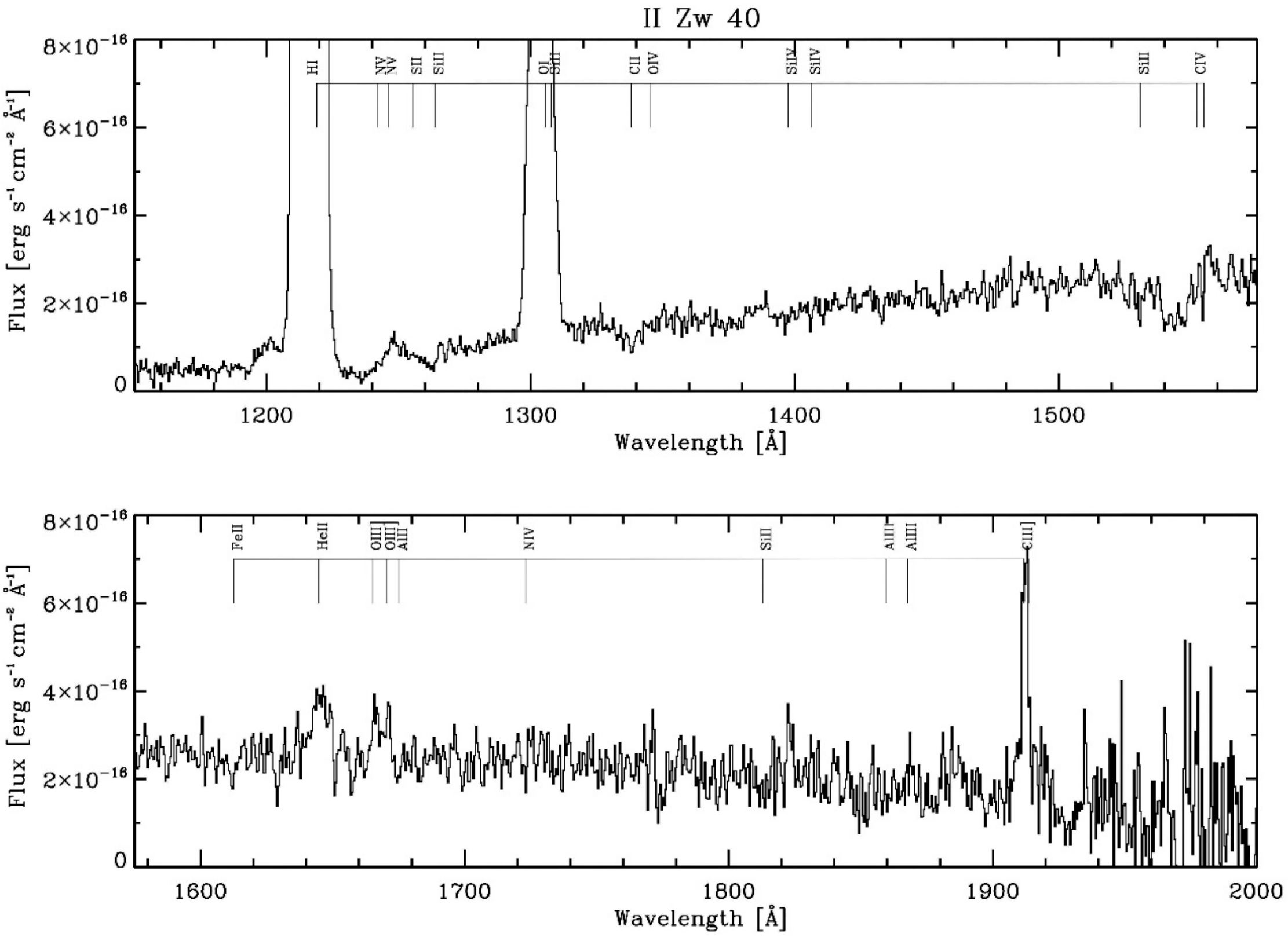

**Figure 17.** Observed UV spectrum of SSC-N. No correction for reddening or redshift was applied. Line identifications are above each spectrum. From [76].

The strongest stellar (in contrast to interstellar or nebular) lines are N V 1240 and C IV 1550, both of which display P Cygni profiles. He II 1640 is the only other clearly detected line with a stellar origin. The width of the He II feature unambiguously indicates a non-nebular origin, although some nebular contribution cannot be excluded. Compared to the N V 1240 and C IV 1550 lines, broad He II 1640 emission is powered by stars that are hotter, have denser winds, and are enriched in helium. These properties are ascribed to W-R stars. The presence of He II 1640 in the UV is consistent with the presence of broad He II 4686 at optical wavelengths by [78]. Star-forming galaxies often show stellar Si IV 1400 (see Figure 10), but this line is absent in the spectrum of SSC-N.

The strongest nebular emission line is C III] 1908, which is often detected as a strong line in mostly metal-poor star-forming galaxies [79]. The only other nebular emission line with a clear detection is O III] 1661,66. Si III] 1883 is detected as well, but at a lower significance. There are several interstellar absorption lines in the spectrum, both foreground and intrinsic. The strongest lines are Si II 1260, C II 1335, and Si II 1526.

The stellar properties of SSC-N can be determined by comparing the observed spectrum with simulated Starburst99 models. These models use stellar evolution models for massive stars accounting for rotation. The evolutionary tracks were linked with a theoretical stellar library of OB star spectra. The spectra were computed with the Wind Model- (WM)-Basic code, which treats stellar winds using spherically extended, expanding, non-LTE atmospheres [80]. W-R stars were accounted for with the Potsdam Wolf-Rayet (PoWR) atmospheres [81]. A single stellar population with a Salpeter-like IMF was assumed.

There are three adjustable parameters in the models: the internal dust reddening $E(B-V)_{\rm int}$, the cluster mass $M$, and the age $T$ of the newly formed massive stars. $E(B-V)_{\rm int}$ is determined from the continuum slope of the spectrum, which otherwise remains unchanged during an O-star dominated phase. The cluster mass is derived from the reddening-free continuum luminosity. The line profiles of the stellar-wind lines provide the age of the population. The oxygen abundance of II Zw 40 is log O/H + 12 = 8.1 [82]. The adopted evolution model with the closest abundance match has log O/H + 12 = 7.9.

In Figure 18, the comparison of the best-fit model (blue spectrum) and the observations is shown. The observed spectrum has been adjusted for foreground reddening. The derived internal reddening of $E(B-V)_{\rm int} = 0.07 \pm 0.03$ is very small; the combined observed reddening is almost entirely caused by the high Galactic foreground reddening.

The derived cluster mass of $(9.1 \pm 1.0) \times 10^5$ $M_\odot$ surpasses that of NGC 2070, the star cluster in the center of 30 Doradus [13] by an order of magnitude. This mass is comparable to that of the optically obscured most massive star cluster in NGC 5253 [83]. It rivals the masses of the most massive young star clusters found in the local universe [84], including those located in the archetypal Antennae galaxies [85]. The mass is similar to values determined for the most massive globular clusters in the Milky Way [86].

An age of $T = (2.8 \pm 0.1)$ Myr is determined from the profiles of the N V, Si IV, and C IV lines, labeled in Figure 18. The strength of N V and C IV decreases with age (see Figure 13), as the most massive stars have died. Si IV is weak early in the evolution of a cluster but increases in strength after about 3 Myr when the first stars leave the main-sequence and become supergiants in the evolution models. The rise in luminosity from dwarf to supergiant stars leads to increased mass-loss rates and wind densities, which then triggers recombination from $Si^{4+}$ to $Si^{3+}$ [87]. This effect is a sensitive age indicator, and the absence of the Si IV 1400 line in SSC-N gives a strong upper age limit. The model fit to the three wind lines of N V, Si IV, and C IV lines is very good. Changes to the IMF would change the theoretical profiles. However, no such adjustment was found to be necessary; the standard IMF results in very good agreement with the observations. It is important to keep in mind that the wind lines are not sensitive to stars with masses of less than ~20 $M_\odot$, and IMF variations below this mass would go unnoticed. The age of SSC-N is similar to the lifetime of stars with masses above 150 $M_\odot$ (~2.8 Myr; [71]). Therefore, this mass regime is unconstrained by the profiles of NV, Si IV, and C IV. For comparison, the model with solar chemical composition in Figure 18 gives a somewhat inferior, but still reasonable fit to the data. The resulting parameters would be $E(B-V)_{\rm int}$ = 0.06 ± 0.03, $M$ = (8.1 ± 1.0) × $10^5$ $M_{\odot}$, and $T$ = (3.0 ± 0.2) Myr, which do not significantly differ from those derived with the metal-poor simulation.

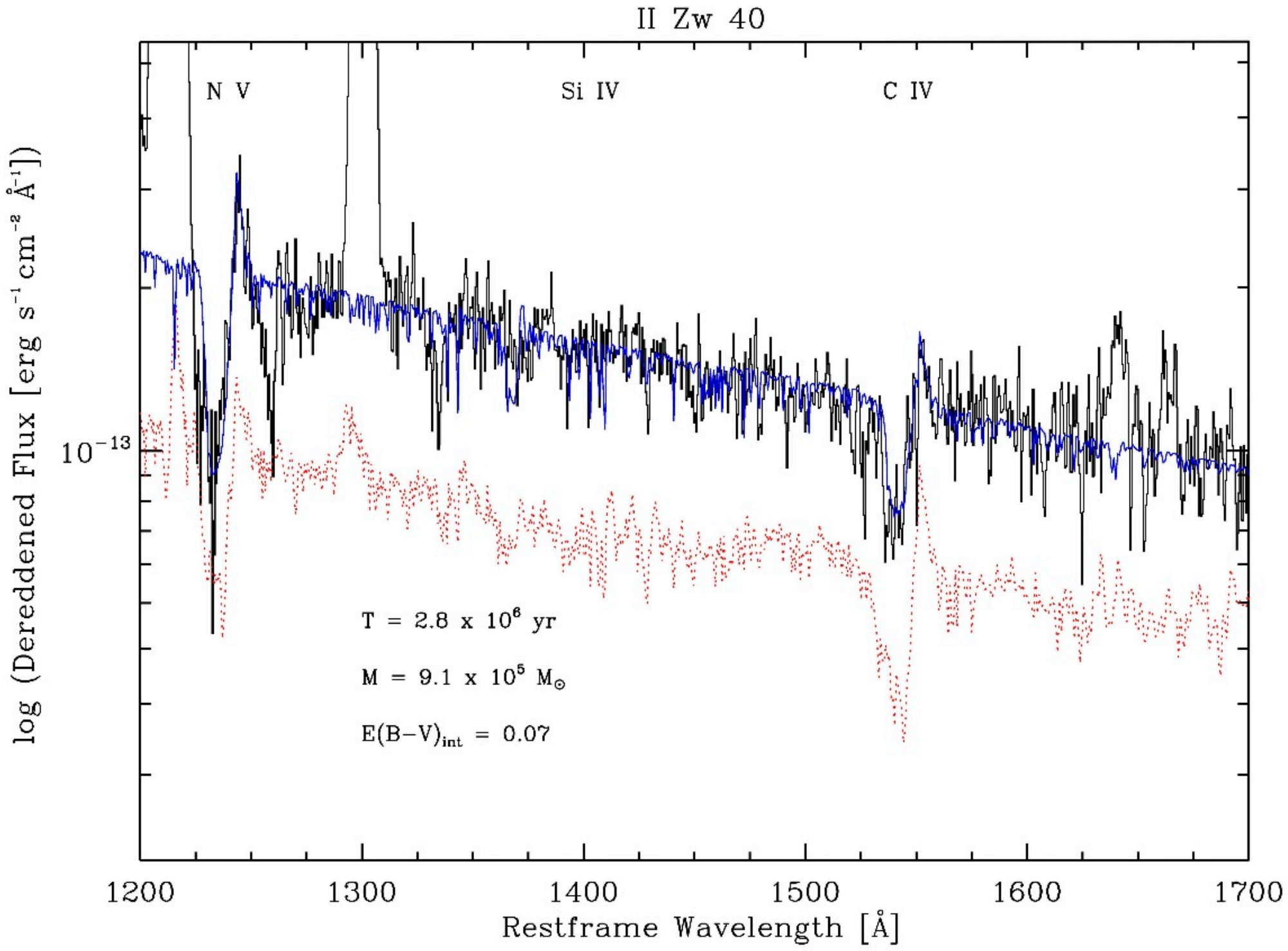

**Figure 18.** Comparison of the observed spectrum of SSC-N (black) and the best-fit simulation (blue). The data are in restframe wavelengths and have been corrected for the Milky Way and internal reddening. The model assumes a star cluster of mass 9.1 × $10^5$ $M_{\odot}$, age 2.8 × $10^6$ yr and chemical composition of 1/7$^{\rm th}$ solar. For comparison, the best-fit model with solar chemical composition is shown as well (red, dashed). From [76].

The derived age of 2.8 Myr agrees with other independent age estimates. In [77], an age of less than ~3.5 Myr from the relative contributions of the thermal and non-thermal radio emission at cm wavelengths was found. If the age were older, core-collapse supernovae (SNe) would enhance the non-thermal emission. A young age and few core-collapse SNe are also indicated by the weak [Fe II] 1.3 and 1.6 µm emission in near-IR spectra of II Zw 40 [88]. These lines are strong when core-collapse SNe have exploded. Iron is strongly depleted by dust in the ISM. Shocks from SN explosions destroy the dust grains, which releases the iron and enhances the [Fe II] lines [89]. The weakness of both lines in SSC-N supports a young age.

Other spectral features seen in Figure 18 are non-stellar and are therefore not included in the simulated spectra—except for He II 1640. The line is usually interpreted as being due to W-R stars, which in principle are included in both the stellar evolution models and in the stellar library. Yet the best-fit model does not produce He II with significant strength at the age of 2.8 Myr because the stars have not yet reached the W-R phase. For a quantitative illustration of the discrepancy, the strength and temporal evolution of He II 1640 EW predicted by Starburst99 is shown in Figure 19. The four lines in the figure correspond to four individual sets of stellar evolution models: models with solar chemical abundance and with 1/7$^{\rm th}$ solar abundance (called subsolar in the figure) with zero rotation and with a rotation velocity of 40% of the break-up speed on the zero-age main-sequence. The subsolar models with 40% break-up velocity were used for the best-fit spectrum in Figure 18. The observed value for SSC-N does not match any of the models. The best agreement is with the rotating solar abundance model at ages of 4 and 8 Myr. This model generates the highest number of W-R stars for several reasons: maximum chemical abundance, maximum rotation velocity, maximum W-R luminosity and maximum mass-loss rates, all of which depend on each other. The secondary bump in this model after 6 Myr is produced by the return of the evolutionary phases from the red supergiant to the W-R domain. However, solar composition is clearly excluded for SSC-N; therefore, these models are not applicable. When the sub-solar models are compared with SSC-N, the disagreement between observations and models worsens. W-R stars never form in significant numbers in the subsolar models for any rotation velocity. They do, in fact, produce hot stars. However, these stars have little or no nitrogen enrichment on the surface and are there not classified as W-R stars, nor would they show W-R features. A possible interpretation would be a more complex star formation history with discrete bursts of star formation and the W-R stars forming earlier. However, the failure of the synthesis models is more likely to be caused by deficiencies in the stellar evolution models. SSC-N is at an early age; therefore, almost all stars are still on the main sequence (in the definition of stellar evolution). The inability of the models to account for the strength of the W-R feature has been found in R136 as well [11]. A plausible explanation is the lack of mixing processes in the evolution models and the subsequent failure to produce chemically enriched stars early-on. Possible mixing processes are convection, mass loss, rotation, or gravitational interaction in binary systems.

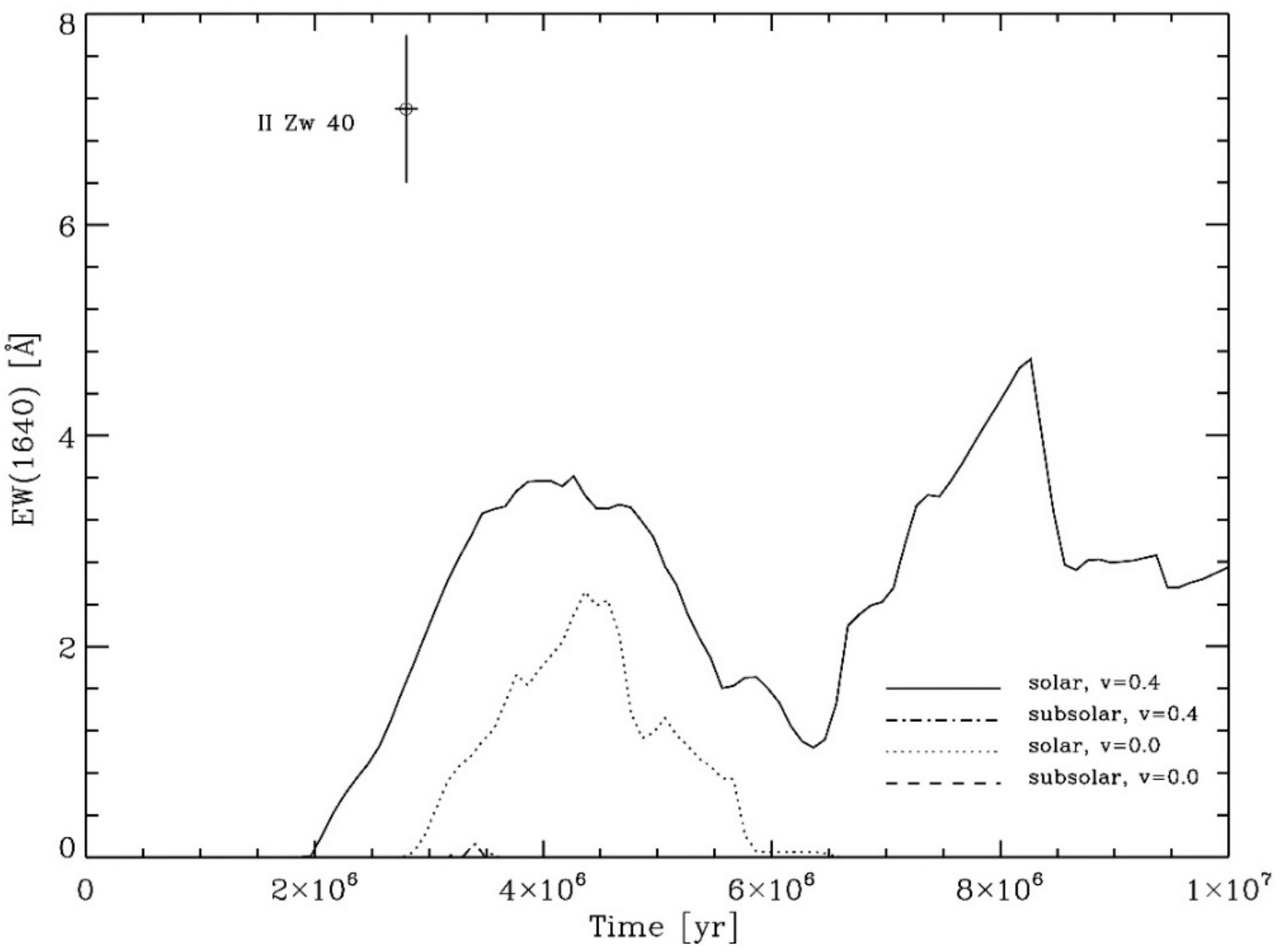

**Figure 19.** Equivalent width of stellar He II 1640 versus time. The value for SSC-N is the open symbol with error bars. The lines show the predictions from four different simulations: solar and subsolar compositions with and without rotation. The predictions for the two subsolar models are very small at any time and are therefore not visible in the graph. From [76].

Published ancillary optical data can be used for gaining additional insight into the properties of SSC-N. Standard optical emission-line ratios determine the location of SSC-N in a standard Baldwin Phillips & Terlevich (BPT) diagram [90] and permit a comparison with related objects (see the left part of Figure 20). The figure includes the locus and density of star-forming galaxies from the *Sloan Digital Sky Survey* (*SDSS*) [91]. Also shown is the sample of metal-poor BCDs of [30], the sample of extreme Green Pea (GP) [92], the local GP analog Mrk 71 [93], and the data for star-forming galaxies at $z \approx 2$–3 of [94]. The location of II Zw 40 coincides with that of the star-forming region in the BPT diagram, which is consistent with the assumption that emission is not powered by an obscured AGN. The 30 Doradus nebula occupies a similar location in the BPT diagram as well [93].

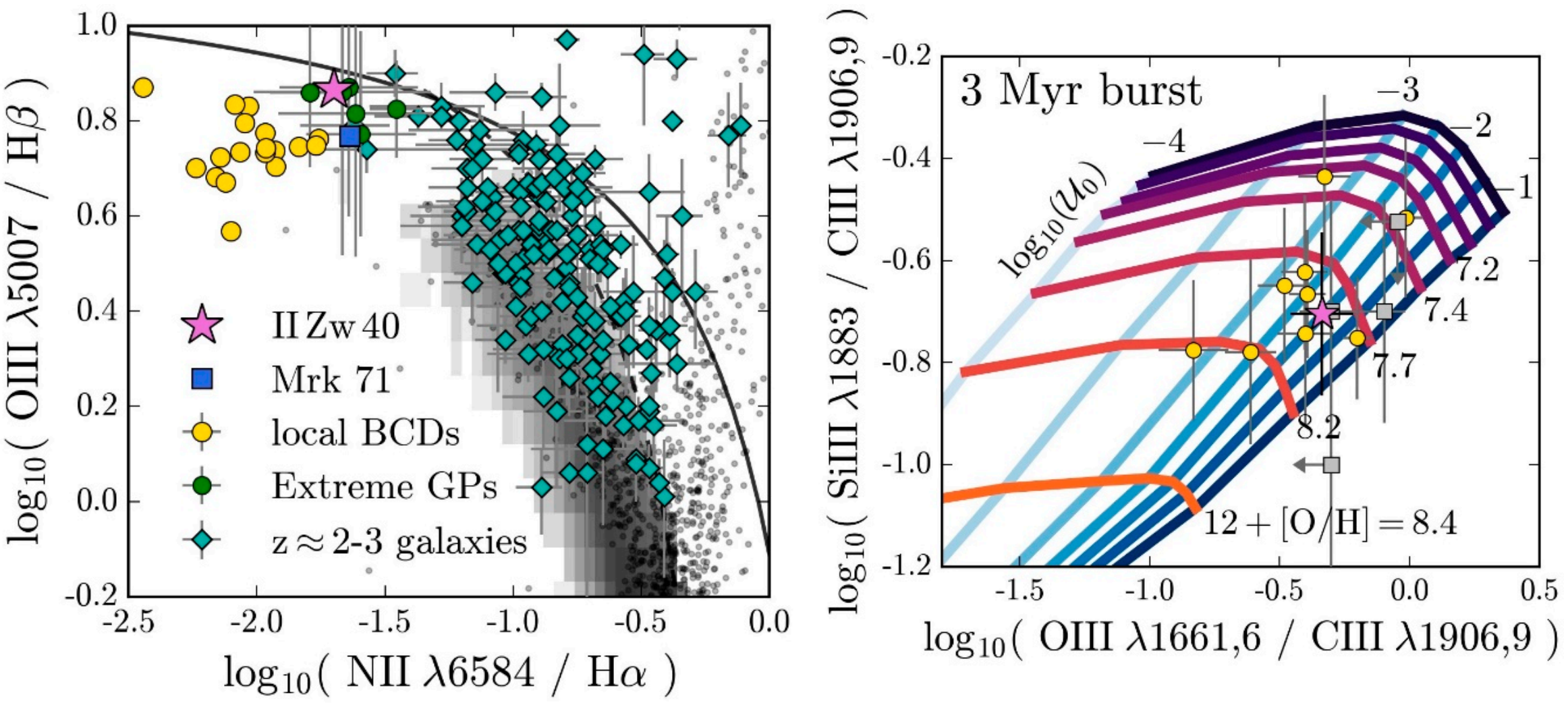

**Figure 20.** Left: Baldwin Phillips & Terlevich (BPT) diagram comparing II Zw 40 to other samples: the BCD sample of [30]; extreme Green Peas of [92]; Mrk 71 [93]; $z \approx$ 2–3 galaxies of [94]. Solid black line: an extreme starburst classification line from [95]. The grayscale 2D histogram indicates the density of star-forming galaxies in SDSS [91]. Right: the computed line ratios of Si III]/C III] versus O III]/C III], using the models of [96]. Blue lines: calculations with a constant ionization parameter, from log $U_0 = -1$ (dark blue) to log $U_0 = -4$ (light blue). Calculations with constant oxygen abundance are connected with lines having different colors, from dark purple to orange. II Zw 40: purple star; BCD sample of [30]: yellow circles; $z \approx 2$ galaxies of [97]: grey squares. From [76].

The sample of [30] contains normal metal-poor, star-forming galaxies whose luminosities, masses and oxygen abundances are similar to those of II Zw 40. In spite of these similarities, SSC-N is shifted towards a larger [O III]/Hβ ratio, indicating more extreme excitation conditions. The locus of SSC-N overlaps with that of the GPs, a class of compact emission-line galaxies found in *SDSS* via their extraordinary [O III] 5007 emission [98]. GPs are assumed to be local counterparts of high-redshift galaxies by means of their high UV luminosity, low oxygen abundance, low dust content, and high specific star-formation rate [99,100]. Mrk 71 has been suggested as the closest ($D$ = 3.4 Mpc) GP analog [93]. SSC-N and Mrk 71 have an almost identical location in the BPT diagram. Like II Zw 40, Mrk 71 hosts one powerful star cluster, which accounts for most of the ionizing photon supply. However, the mass and luminosity of the Mrk 71 star cluster are lower by an order of magnitude than those of SSC-N.

It is worth highlighting the proximity of SSC-N to the sample of [94]. The star-forming galaxies at $z \approx$ 2–3 have extraordinary [O III]/Hβ ratios, which are offset from the location of local star-forming galaxies. The mechanism responsible for these [O III]/Hβ ratios is still under debate but is often interpreted in terms of higher ionization parameters at higher redshift. In comparison with that sample, SSC-N has a similar [O III]/Hβ ratio but, at the same time, the [N II]/Hα ratio is much lower (~0.5 dex). More importantly, the galaxies in the high-redshift sample are much more massive than SSC-N, and their line ratios refer to the entire galaxy. As a result, multiple ionizing sources power the emission lines.

The nebular lines of C III] 1906,09, O III] 1661,66 and Si III] 1883 detected in the UV spectrum offer complementary diagnostics for studying the properties of SSC-N. The right section of Figure 20 reproduces a UV BPT diagram for the Si III] 1883/C III] 1906,09 versus O III] 1661,66/C III] 1906,09 ratios. As in the optical BPT diagram, SSC-N is compared to other galaxies with available UV spectra, including the BCD sample of [30] and the sample dwarf galaxies at $z \approx 2$ studied by [97]. SSC-N occupies a similar area as that of the comparison galaxies, with a high value of the ionization parameter log $U_0$. Superposed on the data is a grid of photo-ionization models of [96], calculated for parameters consistent with those determined for the star-cluster population. A comparison of SSC-N to the model grid implies log $U_0 = -2.0 \pm 0.8$ and log O/H + 12 = 7.99 ± 0.20, which is in excellent agreement with the oxygen abundance of 12 + log O/H = 8.09 derived in the optical. The line ratios in Figure 20 have low ionization energies (< 36 eV) and are therefore not very sensitive to IMF adjustments. An IMF extrapolated to 300 $M_\odot$ leads to essentially the same results for these line ratios in this abundance regime [101]. This no longer holds for lines with relatively high ionization energies such as, e.g., C IV 1550 or N V 1240.

The observed line strengths of C III] 1906,09 and O III] 1661,66 permit a derivation of the ionic abundance ratio of $C^{++}/O^{++}$. Following the method of [102] and [103], $C^{++}/O^{++} = 0.108 \pm 0.012$ is found. Total elemental abundances can be derived for an assumed ionization correction factor (ICF). Photoionization modeling of [104] predicts $C^{++}$ and $O^{++}$ to be the dominant ionization states, resulting in an ICF of order unity. The results of [104] suggest that ICF = 1.1. Adopting the logarithmic depletion factors of −0.30 and −0.07 for C and O of [96] and ICF = 1.1 gives log C/O = −0.70 ± 0.09. This ratio is close to the values of [97] for young low-mass galaxies at $z \approx 2$, as well as for the sample of [30]. In [103], it was determined that log C/O = −0.68 ± 0.13 in ~1000 Lyman-break galaxies at $z \approx 3$.

SSC-N in II Zw 40 is extraordinary in many aspects. The UV spectrum is outstanding in terms of the strong stellar He II 1640 and the nebular O III] 1666 and C III] 1909. The star cluster and the associated H II region surpass the ionizing photon output and stellar mass of the local Rosetta Stone 30 Doradus by an order of magnitude. In the BPT diagram, the SSC-N nebula is offset from the location of local galaxies. Rather, it shares the location of GP galaxies, objects that are often assumed to be nearby analogs of the galaxies capable of reionizing the universe. SSC-N may therefore serve as an invaluable training ground for studying star formation in extreme environments.

## 6. The Future—*ULLYSES* and *CLASSY*

Star formation in the UV will be a major science theme with *HST* for the coming years. Two major surveys will collect hundreds of medium-resolution spectra of individual hot, massive stars in the Local Group of galaxies and of massive star clusters and star forming galaxies out to tens of Mpc.

*Hubble UV Legacy Library of Young Stars as Essential Standards* (*ULLYSES*)[1] will serve as a UV spectroscopic reference sample of high-mass and low-mass, young stars. The library will provide observations that uniformly sample the fundamental astrophysical parameter space, i.e., spectral type, luminosity class, and metallicity. The goal of the Hubble *ULLYSES* library for massive stars is to provide the fundamental reference data set for UV spectroscopy at low metallicity by constructing a comprehensive UV spectral atlas at high spectral resolution ($R > 15{,}000$) using *COS* and *STIS* in the Magellanic Clouds, supplemented by medium-resolution spectroscopy for OB stars in more distant Local Group galaxies. The observations of high-mass stars will consist of 200 orbits on LMC targets, 250 orbits for those in the SMC, and 50 orbits for other low-metallicity Local Group galaxies. This data set will also enable absorption-line studies of the ISM in these galaxies, and the foreground Milky Way, to study element abundances, dust, and multiphase gas kinematics, including galaxy-scale flows. The major enabled science supported by the *ULLYSES* data on young, high-mass stars is as follows.

- *Stellar atmospheres and evolution*: The UV provides access to P Cygni profiles from hot, luminous stars from which wind properties (velocities, mass-loss rates, clumping, porosity) will be empirically obtained. They strongly influence the evolution of massive stars, yet evolutionary calculations often have to rely on theoretical predictions. Furthermore, photospheric lines from carbon, nitrogen, oxygen and the iron forest provide a direct signature of the ionization conditions of iron and other elements within the stellar atmosphere, which are necessary for evaluating line blanketing and mixing. Such information is essential for deriving reliable relations between spectral type and effective temperature, which are in turn necessary for placing stars in the

1 http://www.stsci.edu/stsci-research/research-topics-and-programs/ullyses

Hertzsprung-Russell diagram and understanding their evolution. The high-resolution spectra will also yield the projected rotational velocities, which are another vital parameter affecting stellar evolution and Lyman continuum luminosities.

- *Spectral templates for stellar population synthesis*: The library will provide the much-needed OB and W-R spectroscopic templates for rest-frame UV studies of integrated stellar populations in high-$z$ galaxies with the James Webb Space Telescope and Extremely Large Telescopes. The proximity and low metallicity of the LMC and SMC makes them ideal targets. The atlas will greatly extend the number of high-quality, UV spectroscopic templates in both galaxies, achieving a similar OB and W-R sample to that of the Milky Way from *IUE*. It will also provide more representative examples, since archival datasets were largely selected based on other criteria, such as being UV-bright for ISM studies, or focused on unusual systems (e.g., magnetic O stars, rapid rotators). Currently, low-metallicity templates are poorly sampled compared to those at solar values, yet the former are essential for interpreting the stellar populations in starburst galaxies such as GPs and Lyman-$\alpha$ emitters.
- *Stellar populations at low metallicity*: The spectral templates will clarify the IMF, cluster ages, and ionizing SED in massive clusters and local galaxies that serve as analogs of higher redshift objects, which are commonly metal-deficient, especially in iron peak elements. These are critical for estimating star cluster masses and ages, which are the fundamental input parameters for understanding massive-star feedback and evolutionary processes in star-forming galaxies. Indeed, some local, intensely star-forming, metal-poor galaxies are known to be Lyman continuum emitters; identifying their stellar populations is critical to understand the conditions for Lyman continuum escape. In addition, stellar abundances at low metallicity are more accurate in weak-wind populations, and can calibrate nebular diagnostics.
- *Multi-phase ISM and dust*: The stellar spectra will contain many interstellar metal lines across the UV. This will enable comprehensive studies of the ISM in the Magellanic Clouds, Milky Way, and perhaps the metal-poor galaxies; in particular, element abundances, dust depletion, kinematics, ionization state, and spatial distribution of multi-phase gas. UV continuum studies will further characterize the dust extinction law in a range of metallicities and environments, since the foreground Milky Way component of the extinction is low.
- *Circumgalactic medium*: The stellar spectra will also reveal absorption lines from the circumgalactic medium of the Magellanic Clouds and the Milky Way. The LMC systemic velocity of +260 km s$^{-1}$ is large enough to differentiate LMC and Galactic components, while the SMC systemic velocity of +150 km s$^{-1}$ allows probing a more limited velocity range. This data set can thus be leveraged to study the galaxy-scale gas inflows and outflows, clarifying the baryon and metal cycle of star formation, feedback, galactic chemical evolution, and other evolutionary processes in a system that is currently being dynamically entrained by the Milky Way. Moreover, the high sensitivity of UV wavelengths to small particles and large molecules will enable measuring variations in the particles size distribution, and the connection between the abundance of polycyclic aromatic hydrocarbon and UV irradiation.

The *COS Legacy Archive Spectroscopic SurveY* (*CLASSY*)[2] Treasury is a program that builds upon archival data to create the first high-quality, high-resolution *COS* M-mode UV spectral catalog of star-forming galaxies in the local universe using 135 *HST* orbits. The sample of 46 star-forming galaxies was selected to mimic the similar properties observed at high-$z$, with a broad range of chemical abundances, ionization parameters, densities, masses, and star-formation rates. The spectra will be sensitive to key emission and absorption lines from massive stars and the ISM. These spectra can be used to study the massive stellar populations in metal-poor galaxies, the physical properties of

2 https://www.danielleaberg.com/classy

powerful outflows that regulate star formation, and the chemical abundance characteristics of the gas and stars. *CLASSY* will enhance the diagnostic power of the UV lines for upcoming *JWST/ELT* surveys, offering a lasting legacy to the community.

Combining G130M+G160M+G185M spectra will provide observations of many emission and absorption lines that are important for characterizing the ionizing stellar population and physical conditions of the nebular gas. The main objective of this program is to unify the stellar and gas-phase physics, allowing a holistic understanding of massive stars as the drivers of the gaseous evolution of star-forming galaxies. The scientific objectives include:

- *The effects of massive stars on the surrounding gas*: The radiation emitted by massive stars influences all aspects of UV spectra in star-forming galaxies, yet their ionizing spectra are not well understood. They determine the shape of the far-UV continuum and their extreme-UV (EUV) radiation fields are reprocessed by the ISM, powering the nebular continuum and emission lines. Uncertainties in the shape of the ionizing spectrum significantly affect the interpretation of UV spectra, including gas properties, stellar feedback, production of H-ionizing photons, and effects of dust. While the implementation of new ingredients in stellar population synthesis, such as rotation or binaries, continues to refine the predicted EUV radiation field, the shape of the ionizing spectrum remains very poorly constrained for the metal-poor stellar populations that come to dominate at high redshift.
- *Revealing the physical properties of outflows*: The kinematics of the galaxy-scale outflows of gas driven by massive stars are encoded into the Lyman-$\alpha$ profiles and ISM resonant absorption lines observed in the UV. Since these outflows are likely photoionized, the observed stellar population properties help to determine the properties of the outflowing gas. In turn, the ISM absorption lines from multiple ions spanning the UV coverage will determine the outflow's ionization structure, chemical composition, and gas mass. These measurements are important for determining the total gas mass removed by stellar feedback and constraining the energy injected by the observed massive-star population, and are uniquely probed in the UV.
- *UV diagnostics of chemical evolution*: Studying the chemistry and physical conditions in star-forming galaxies is key to understanding the principal components of galaxy formation and evolution: outflows, infall, star-formation, and gas enrichment. Traditional optical emission-line diagnostics used to investigate such properties (i.e., the metal content, density, and the strength/shape of ionizing radiation) will not be accessible for the most distant galaxies observed with *ELT*s and *JWST*, highlighting the need for well-calibrated tracers at UV wavelengths. Strong UV lines characterize a plethora of gas properties, including temperature, density, and metal abundance, as well as reflecting the properties of the ionizing spectrum.
- *Exploring reionization physics*: At redshifts between $z$ = 6–10, ionizing photons escaped from galaxies to reionize the universe. Determining the sources of cosmic reionization is one of four key science goals of *JWST*. However, neither *JWST* nor *ELT*s will directly observe the Lyman continuum during the epoch of reionization owing to the increasing opacity of the intergalactic medium. To discern whether star-forming galaxies reionized the universe, indirect indicators must be used to measure: (1) the intrinsic number of ionizing photons produced by massive stars and (2) the fraction of these photons that escape galaxies. The product of these two quantities is the number of ionizing photons emitted by a star-forming galaxy. *CLASSY* will predict the number of ionizing photons from the massive star features and determine correlations with UV emission lines. Theoretical arguments and small observational samples suggest that UV nebular emission and absorption features trace the escape fraction. *CLASSY* will indirectly infer escape fractions of a statistically significant sample using UV diagnostics accessible by *ELT*s and *JWST*: Lyman-$\alpha$ emission, the depth of low-ionization absorption lines, and the strength of high-ionization emission lines.

**Acknowledgments:** Danielle Berg (Principal Investigator—*CLASSY*) kindly provided unpublished information on the *CLASSY* program.

**Conflicts of Interest:** The author declares no conflict of interest.